%% file: DY_qTcorrection.tex
\documentclass[a4paper]{article}

\usepackage{jheppub} 
                     
\usepackage{graphicx,color}
 \usepackage{bm}
   \usepackage{amsmath}
    \usepackage{amssymb}    
     \usepackage{pifont}
 \usepackage{stmaryrd}

\usepackage{standalone}
\usepackage{slashed}
\usepackage{multirow}
\usepackage{makecell}
\usepackage{tabu}
\usepackage{hhline}
\usepackage{colortbl}

\usepackage{tikz}
\usetikzlibrary{decorations.pathmorphing}
\usetikzlibrary{decorations.markings}
\usetikzlibrary{patterns}
\usetikzlibrary{calc}
\usetikzlibrary{math}

\newcommand{\Ds}{\displaystyle}

\newcommand{\nn}{\nonumber}

\newcommand{\Tr}{\mathrm{Tr}}

\renewcommand{\(}{\left(}
\renewcommand{\)}{\right)}
\renewcommand{\[}{\left[}
\renewcommand{\]}{\right]}

\renewcommand{\vec}[1]{\bm{#1}}

\newcommand{\colorThree}[1]{{\color[rgb]{0.8,0.3,0.} #1}}

\title{Leading $q_T/Q$ correction for Drell-Yan  in TMD factorization}

\author{Arturo Arroyo-Castro\footnote{https://orcid.org/0009-0002-6091-1829},}

\author{Ignazio Scimemi\footnote{https://orcid.org/0000-0001-5598-5810},}

\author{Alexey Vladimirov\footnote{https://orcid.org/0000-0001-5449-194X}}

\affiliation{Departamento de F\'isica Te\'orica \& IPARCOS, Universidad Complutense de Madrid, E-28040 Madrid, Spain}

\emailAdd{arturarr@ucm.es}
\emailAdd{ignazios@ucm.es}
\emailAdd{alexeyvl@ucm.es}

\preprint{IPARCOS-UCM-25-021}

\abstract{
The transverse momentum dependent (TMD) factorization theorem accommodates various types of power corrections. Among them, the least studied are $q_T/Q$ corrections, which become significant at large values of transverse momentum. These corrections partially originate from higher-twist TMD distributions, which exhibit singularity at small transverse distances. We propose a decomposition that reveals this singularity explicitly, and makes the $q_T/Q$ correction manifest. As a concrete application, we consider the next-to-leading power correction for the angular distributions in Drell-Yan, and determine the leading $q_T/Q$ corrections. These corrections are significant for the angular distributions $A_1$ and $A_3$, in complete agreement with the data.
}

\begin{document} 
\allowdisplaybreaks
\maketitle 

\section{Introduction}

The transverse momentum dependent (TMD) factorization theorem is a result of an expansion  of the cross-section in powers of a hard scale~\cite{Collins:2011zzd, Echevarria:2011epo, Becher:2010tm} and it plays a crucial role in describing collision processes and understanding the structure of hadrons. The leading power (LP) term of the TMD factorization approach successfully describes a wide range of observables across various energy scales \cite{Angeles-Martinez:2015sea, Boussarie:2023izj} and it incorporates several twist-two TMD distributions, which are extracted from experimental data (for the most recent determinations, see refs.~\cite{Moos:2025sal, Bacchetta:2025ara, Bacchetta:2024qre, Bacchetta:2024yzl, Yang:2024drd, Billis:2024dqq}). Nevertheless, the LP approximation has its limitations, mainly due to the constraint $q_T\ll Q$, where $q_T$ is the transverse momentum of the probing particle and $Q$ is its virtuality, so that power corrections must be added to improve our comprehension. In this work, we investigate the structure of the next-to-leading power (NLP) term in TMD factorization and identify the contributions that govern the leading $q_T$ behavior of the cross-section.

Studies on power corrections to the TMD factorization theorem have advanced significantly in recent years. There are several computations at the next-to-leading power (NLP) and at the next-to-next-to-leading power (NNLP)~\cite{Balitsky:2017gis, Balitsky:2020jzt, Inglis-Whalen:2021bea, Balitsky:2021fer, Vladimirov:2021hdn, Ebert:2021jhy, Rodini:2022wic, Gamberg:2022lju, Rodini:2023plb, delCastillo:2023rng, Balitsky:2024ozy}. Herewith, the NLP term is suppressed by $1/Q$, the NNLP term by $1/Q^2$, and so forth.

It is important to remark that the expansion in inverse powers of $Q$ is not the only defining characteristic of a power correction term. The structure of power corrections in TMD factorization is very involved due to the multi-dimensional nature of both observables and partonic distributions. The factors of $1/Q$ can be accompanied by various kinematic scales, such as $q_T$ (the transverse momentum of the probe), $k_T$ (the transverse momentum of the parton), $M$ (the hadron mass), or $\Lambda$ (a generic hadronic scale). As a result, different types of power corrections become dominant in specific kinematic regimes.

The factorized cross-section at NLP consists of two principal contributions
\begin{eqnarray}\label{intro1}
d\sigma^{\text{NLP}}=d\sigma^{\text{KPC}}+d\sigma^{\text{gNLP}},
\end{eqnarray}
where $d\sigma^{\text{KPC}}$ is the contribution of kinematic power corrections (KPC), and $d\sigma^{\text{gNLP}}$ is the contribution of genuine power corrections (gNLP). The terms with KPC are contain derivatives of TMD distributions of twist-two, which introduce an explicit dependence on the (partonic) transverse momentum $k_T$, and thus are named $k_T/Q$ corrections. Arguably, KPC are the most significant, since they are responsible for the restoration of frame-invariance and transversality of the hadron tensor. A detailed investigation of KPC is provided in ref.~\cite{Vladimirov:2023aot}. Genuine power corrections incorporate TMD distributions of twist-three, and therefore are dubbed as $\Lambda/Q$ corrections. Theoretical properties of twist-three TMD distributions are known~\cite{Rodini:2022wki}, but currently only a little practical information is available about them.

The decomposition in eq.~(\ref{intro1}) follows directly from the operator analysis  \cite{Balitsky:2020jzt, Vladimirov:2021hdn}, and it clearly misses $q_T/Q$ terms, which form of the large-$q_T$ asymptotic behavior (the so-called $Y$-term \cite{Collins:2011zzd}). One can guess that such corrections appear only at NNLP, where they are generated by the real-particles-exchange diagrams. However, this is not the only source of such corrections. As we demonstrate in this work, a $q_T/Q$ contribution is  hidden in the genuine power correction term.

The origin of the $q_T/Q$ behavior lies in the twist-three TMD distributions themselves. Higher-twist TMD distributions are singular in the limit $b\to0$ (where $b$ is the Fourier conjugate of $k_T$) \cite{Rodini:2022wki}. In momentum space this implies that they exhibit power-like growth as $k_T\to\infty$. This singularity generates terms that scale as $q_T/Q$, instead of the expected $\Lambda/Q$, thereby spoiling a straightforward power counting.

We propose here a decomposition/redefinition of twist-three TMD distributions\footnote{The concept can be generalized to higher-twist cases as well.} that explicitly reveals this singularity and parametrizes it in terms of twist-two TMD distributions. We construct
\begin{eqnarray}\label{intro3}
\Phi_3=\widehat{\Phi}_3+S\otimes \Phi_2,
\end{eqnarray}
where $\Phi_{3(2)}$ is a TMD distribution of twist-three(-two), $S$ is a perturbative kernel that contains the singularity and $\otimes$ is an integral convolution. The function $\widehat{\Phi}_3$ is regular. Applying this decomposition, the cross-section, eq.~(\ref{intro1}), turns into
\begin{eqnarray}\label{intro2}
d\sigma^{\text{NLP}}=d\sigma^{\text{KPC}}+\widehat{d\sigma}^{\text{gNLP}}+\frac{q_T}{Q}d\sigma^{\text{L-TMD}},
\end{eqnarray}
where $\widehat{d\sigma}^{\text{gNLP}}$ represents $d\sigma^{\text{gNLP}}$ with all twist-three distributions replaced by their regular parts $\widehat{\Phi}_3$, and $d\sigma^{\text{L-TMD}}$ contains only distributions of twist-two. The label L-TMD refers to the term ``leading TMD approximation'', which we use here to refer to the terms $S\otimes \Phi_2$, as these terms describe the dominant behavior of TMD distributions at large transverse momentum. In eq.~(\ref{intro2}), one can easily identify the power corrections of the type $k_T/Q$, $\Lambda/Q$ and $q_T/Q$. Moreover, at sufficiently large-$Q$ (that is $k_T/Q,\;q_T/Q\gg \Lambda/Q$), the term $\widehat{d\sigma}^{\text{gNLP}}$ can be neglected, while KPC and L-TMD terms remain and provide a good approximation for observables.

In this work, we elaborate this scheme and we show an explicit case. For concreteness, we consider the angular distribution of the Drell-Yan lepton pair produced by a neutral boson. This observable serves as an ideal testing ground for studying of power corrections, since four angular distributions are generated at NLP. Moreover, these distributions have been measured at the LHC \cite{ATLAS:2016rnf, CMS:2015cyj, LHCb:2022tbc} with sufficient precision to allow a meaningful comparison with theoretical expectations.

Angular distributions have been extensively studied from the perspective of collinear factorization (i.e., at large and moderate $q_T$), see the most recent works \cite{Karlberg:2014qua, Lambertsen:2016wgj, Gauld:2017tww, Gauld:2021pkr, Lyubovitskij:2024civ, Lyubovitskij:2024jlb, Lyubovitskij:2025oig}. In contrast, there have been only a few studies within the framework of TMD factorization approach (i.e., at low $q_T$) \cite{Barone:2010gk, Lu:2011mz, Ebert:2020dfc, Balitsky:2021fer, Piloneta:2024aac}. In ref.~\cite{Piloneta:2024aac}, it was shown that (resummed) KPCs provide a good description of many angular distributions (the ones with sufficiently precise measurements), with a notable exception of the $A_1$ distribution, where the prediction is approximately twice as large as the measurement. In this work, we show that this discrepancy arises from the $q_T/Q$ correction. After addition of the $q_T/Q$ correction, the theoretical prediction agrees with the data.

This paper is structured as follows. In the sec.~\ref{sec:NLP}, we present the angular distributions derived from the TMD factorization theorem at NLP. To derive it, we use the known results for the hadron tensor taken from ref.~\cite{Vladimirov:2021hdn}, and presented in a re-structured from in sec.~\ref{sec:HT}. The LP and NLP  for angular distributions are presented in sec.~\ref{sec:AS}. The derivation of the L-TMD approximation is done in sec.~\ref{sec:L-TMD}. We start from the general discussion of its construction in sec.~\ref{sec:L-TMD-general}, and proceed with derivation of TMD distributions at LO in sections \ref{sec:L-TMD-LO} and \ref{sec:L-TMD-LA}. In sec.~\ref{sec:L-TMD-AS}, we apply this approximation to the angular coefficients, derive the $q_T/Q$ correction and compare the result with the data. For completeness, we collect the lepton tensor and the details about electro-weak structure in appendices \ref{app:lepton-tensor} and \ref{app:couplings}.

\section{TMD factorization for unpolarized DY at NLP}
\label{sec:NLP}

We consider the reaction of lepton pair production in the hadron-hadron collision
\begin{eqnarray}\label{reaction}
h_1(p_1)+h_2(p_2)~\to~ \gamma^*/Z(q)+X ~\to~ \ell^-(l)+\ell^+(l') +X,
\end{eqnarray}
 where the momenta of the respective particles are indicated in brackets. For simplicity, we assume that the leptons and hadrons are massless, $l^2=l'^2=p_{1,2}^2=0$, and unpolarized. The cross-section of the Drell-Yan reaction has a non-trivial dependence on the helicity of vector boson, which can be extracted from the azimuthal modulations of the lepton pair. The standard parametrization of the angular structure is defined in the Collins-Soper frame \cite{Collins:1977iv} and is given by \cite{Mirkes:1994eb}
\begin{eqnarray}\label{def:angular}
\frac{d\sigma}{d^4qd\Omega}&=&\frac{3}{16\pi} \frac{d\sigma}{d^4q}\Big[
(1+\cos^2\theta)+\frac{1-3\cos^2\theta}{2}A_0+\sin2\theta \cos \phi A_1
+\frac{\sin^2\theta \cos2\phi}{2}A_2
\\\nn && 
+\sin\theta \cos\phi A_3+\cos\theta A_4
+\sin^2\theta \sin 2\phi A_5+\sin2\theta \sin\phi A_6+\sin\theta\sin\phi A_7\Big]
\\ \nn &=&\frac{3}{16\pi} \frac{d\sigma}{d^4q}\sum_{n=U,0,...,7}S_n(\theta,\phi)A_n,
\end{eqnarray}
where $A_n$ are the angular distributions. Angular distributions $A_i$ exhibit a variety of properties and structures with respect to electroweak and strong interaction (see, for instance, the synopsis in ref.~\cite{Piloneta:2024aac, Lyubovitskij:2024civ}). In what follows, we derive the angular distributions in  TMD factorization theorem at NLP.

The differential cross-section of the DY reaction, eq.~(\ref{reaction}), expressed as a convolution of lepton and hadron tensors is
\begin{eqnarray}\label{dsigma:0}
d\sigma=\frac{2\alpha_{\text{em}}^2}{s} \frac{d^3l}{2E}\frac{d^3l'}{2E'}\sum_{GG'}L_{GG',\,\mu\nu}
W_{GG'}^{\mu\nu}\Delta^*_G(q)\Delta_{G'}(q).
\end{eqnarray}
Here, $\alpha_{\text{em}}=e^2/4\pi$ is the QED coupling constant, $s=(p_1+p_2)^2$ is the center-of-mass energy, the index $G$ runs over the gauge boson type, and $\Delta_G$ is the propagator of the corresponding gauge boson. Note that although we consider neutral boson production, the results can be easily extended to $W$-boson production by appropriately modifying the coupling constants and propagators in the final formulas.

The dependence on the azimuthal angles of the lepton pair $\theta$ and $\phi$ is entirely contained in the lepton tensor, which can be written as
\begin{eqnarray}\label{L-tensor}
L^{\mu\nu}_{GG'}&=&(-Q^2)\(
z^{GG'}_{+\ell}\sum_{n=U,0,1,2,5,6}S_n(\theta,\phi)\mathfrak{L}_n^{\mu\nu}
+
z^{GG'}_{-\ell}\sum_{n=3,4,7}S_n(\theta,\phi)\mathfrak{L}_n^{\mu\nu}\),
\end{eqnarray}
where $S_n$ are angular modulations that correspond to distributions $A_n$ (they are given explicitly in eq.~(\ref{S-def})), and 
\begin{eqnarray}
z^{GG'}_{\pm \ell}&=&2(g^R_Gg^R_{G'}\pm g^L_Gg^L_{G'})
\end{eqnarray}
is the combination of electro-weak coupling constants for the lepton. The tensors $\mathfrak{L}_n^{\mu\nu}$ can be expressed in  terms of the vectors $\bar n$, $n$ and $q$, i.e. using the  hadronic momenta, which greatly simplifies the algebraic computation. The expressions for $\mathfrak{L}^{\mu\nu}_n$  were found in ref.~\cite{Piloneta:2024aac}, and are summarized in the appendix \ref{app:lepton-tensor}. Combining together the cross-section in eq.~(\ref{dsigma:0}) and the lepton tensor in eq.~(\ref{L-tensor}), one finds  the coefficients of the angular decomposition in eq.~(\ref{def:angular}),
\begin{eqnarray}
\Sigma_U=\frac{d\sigma}{d^4q},\qquad A_n=\frac{\Sigma_n}{\Sigma_U}.
\end{eqnarray}
with
\begin{eqnarray}\label{def:Sigma_n}
\Sigma_n=\frac{4\pi \alpha_{\text{em}}^2}{3s}(-Q^2)\sum_{GG'}\mathfrak{L}_n^{\mu\nu} W^{\mu\nu}_{GG'} z_{n\ell}^{GG'}\Delta_G^*\Delta_G'.
\end{eqnarray}

Next we derive the  angular coefficients in  TMD factorization, performing the following steps:
\begin{enumerate}
\item Computing the hadron tensor in TMD factorization. 
\item Expressing the hadron tensor in terms of TMD distributions by substituting the parametrization of the TMD matrix elements.
\item Contracting hadron and lepton tensors, and integrate over the angular part.
\end{enumerate}
The first step has already been solved in refs.~\cite{Balitsky:2020jzt, Vladimirov:2021hdn}, where one can find the NLP TMD factorization theorem in terms of operators, along with the corresponding coefficient functions at NLO \cite{Vladimirov:2021hdn}. The parametrization of matrix elements is also established; at the twist-three level, it has been developed in ref.~\cite{Rodini:2022wki}. Thus, our task is to carry out the final step, which is merely algebraic.

In the following section, we summarize the relevant results from \cite{Vladimirov:2021hdn, Rodini:2022wki, Rodini:2023plb}, which serve as the inputs for the first two steps. Then in the section \ref{sec:AS} we present the result for contraction of tensors and the $\Sigma_n$ at  LP and NLP.

\subsection{Hadron tensor}
\label{sec:HT}

The hadron tensor for DY  is defined as
\begin{eqnarray}\label{def:W}
W_{GG'}^{\mu\nu}&=&\int \frac{d^4y}{(2\pi)^4}e^{-i(qy)}\langle p_1,p_2|J_G^{\dagger \mu}(y)J_{G'}^\nu(0)|p_1,p_2\rangle,
\end{eqnarray}
where $J_G^\mu$ is the electro-weak current for the gauge boson $G$. It reads
\begin{eqnarray}
J_G^\mu=\bar q(g_R^G\gamma^\mu (1+\gamma^5)+g_L^G(1-\gamma^5))q,
\end{eqnarray}
where $g_{R,L}$ are the left and right coupling constants for the gauge boson. For future convenience, we introduce the shorthand notation for the Dirac structure of the current
\begin{eqnarray}
\omega_G^\mu=g_R^G\gamma^\mu (1+\gamma^5)+g_L^G(1-\gamma^5),
\end{eqnarray}
such that $J_G^\mu=\bar q\omega_G^\mu q$. For the case of photon production one has $\omega_\gamma^\mu=\gamma^\mu$.

The hadron tensor has three terms  up to the order of our interest,
\begin{eqnarray}\label{WNLP}
W^{\mu\nu}&=&\frac{1}{N_c}\int \frac{d^2b}{(2\pi)^2}e^{-i(qb)}\(\widetilde{W}_{\text{LP}}^{\mu\nu}+\widetilde{W}_{\text{KPC}}^{\mu\nu}+\widetilde{W}_{\text{gNLP}}^{\mu\nu}+...\),
\end{eqnarray}
consisting in a  LP term, a kinematic power correction at NLP, and a genuine power correction at NLP, correspondingly. The dots represent the higher power contributions, that we neglect. In the following sub-sections, we present each term in eq.~(\ref{WNLP}) one-by-one explicitly.

\subsubsection{Leading power term}

The LP term can be written as
\begin{eqnarray}\label{WLP}
\widetilde{W}^{\mu\nu}_{\text{LP}}&=&
\mathbb{C}_{\text{LP}}\(\frac{\tau^2}{\mu^2}\)\sum_{n,m}
\Big[
\frac{1}{4}\Tr\(\omega^\mu_{G} \overline{\Gamma}_m^-\omega^\nu_{G'} \overline{\Gamma}_n^+\)
\Phi_{11}^{[\Gamma^+_n]}(x_1,b;\mu,\zeta) 
\overline{\Phi}_{11}^{[\Gamma_m^-]}(x_2,b;\mu,\bar \zeta)
\\\nn &&
\phantom{\mathbb{C}_{\text{LP}}\sum_{n,m}\Big[}
+
\frac{1}{4}\Tr\(\omega^\mu_{G} \overline{\Gamma}_n^+\omega^\nu_{G'} \overline{\Gamma}_m^-\)
\overline{\Phi}_{11}^{[\Gamma^+_n]}(x_1,b;\mu,\zeta)
\Phi_{11}^{[\Gamma_m^-]}(x_2,b;\mu,\bar \zeta)
\Big],
\end{eqnarray}
where $\Phi_{11}$ and $\overline{\Phi}_{11}$ are quark and anti-quark TMD distributions of twist-two (defined below in eq.~(\ref{def:PHI11}-\ref{def:PHI11-gluon})), $\mathbb{C}_{\text{LP}}$ is the hard coefficient function (given in eq.~(\ref{def:CLP})). The variables $\tau^2$, $x_1$ and $x_2$ are \footnote{We use the standard notation for the light-cone components of the vector
$$
v^\mu=\bar n^\mu v^++n^\mu v^-+v_T^\mu,
$$
where $v_T$ is the transverse part, and $n$ and $\bar n$ are light-cone vectors $n^2=\bar n^2$ with $(n\bar n)=1$. Vectors $n$ and $\bar n$ are associated with the hadron momenta $p_1^\mu=\bar n^\mu p_1^+$ and $p_2^\mu=n^\mu p_2^+$. The scalar product of transverse components written in the bold font is Euclidian, i.e. $\vec q_T^2=|q_T^2|=-q_T^2>0$ and $\vec b^2=|b^2|=-b^2>0$. We also use the standard notation for two-index transverse tensors: $g_T^{\mu\nu}=g^{\mu\nu}-n^\mu \bar n^\nu-\bar n^\mu n^\nu$ and $\epsilon_T^{\mu\nu}=\epsilon^{-+\mu\nu}$.
}
\begin{eqnarray}\label{def:x12}
\tau^2=2q^+q^-=Q^2+\vec q_T^2,\qquad
x_1=\frac{q^+}{p_1^+},\qquad x_2=\frac{q^-}{p_2^-}
\end{eqnarray}
The pairs $\{\Gamma, \overline{\Gamma}\}$ are the elements of the complete Dirac basis that project only good components of the quark spinors. They are
\begin{eqnarray}
\{\Gamma^+, \overline{\Gamma}^+\}:&\qquad&
\{\gamma^+, \gamma^-\},\quad 
\{\gamma^+\gamma^5, -\gamma^-\gamma^5\},\quad 
\{i\sigma^{\alpha +}\gamma^5, -i\sigma^{\alpha -}\gamma^5\},
\\\nn 
\{\Gamma^-, \overline{\Gamma}^-\}:&\qquad&
\{\gamma^-, \gamma^+\},\quad 
\{\gamma^-\gamma^5, -\gamma^+\gamma^5\},\quad 
\{i\sigma^{\alpha -}\gamma^5, -i\sigma^{\alpha +}\gamma^5\}.
\end{eqnarray}
In eq.~(\ref{WLP}), the summation over $n$ and $m$ runs through these pairs. Note that, for brevity, our notation does not explicitly specify the hadrons to which the TMD distributions belong. This information can be inferred from the variables $x_{1,2}$, i.e. a TMD distribution that depends on $x_{1(2)}$ is originated from hadron $h_{1(2)}$. Similarly, we omit the explicit flavor indices, which are understood to be summed over all active quark flavors.

The variables $\zeta$ and $\bar \zeta$ are the scales associated with the renormalization of rapidity divergences. These scales must satisfy 
\begin{eqnarray}\label{zeta*zeta=tau2}
\zeta\bar \zeta =\tau^2=2q^+q^-.
\end{eqnarray}
The scale $\mu$ is the hard factorization scale, and its dependence  cancels in-between the coefficient function and the TMD distributions. The scale dependence of the TMD distributions are dictated by corresponding evolution equations \cite{Aybat:2011zv, Chiu:2012ir, Scimemi:2018xaf}. The coefficient function $\mathbb{C}_{\text{LP}}$ is the LP hard coefficient function. At NLO it reads
\begin{eqnarray}\label{def:CLP}
\mathbb{C}_{\text{LP}}\(\frac{\tau^2}{\mu^2}\)=1+2a_s(\mu)C_F\(-\mathbf{L}^2+3\mathbf{L}-8+\frac{7\pi^2}{6}\)+\mathcal{O}(a_s^2),
\end{eqnarray}
where $C_F=(N_c^2-1)/2N_c$ (for $SU(N_c)$ gauge group), $a_s=g^2/(4\pi)^2$ is the QCD coupling constant, and $\mathbf{L}=\ln(\tau^2/\mu^2)$. Nowadays,  $\mathbb{C}_{\text{LP}}$, as well as, the TMD evolution function are known up to N$^4$LO \cite{Lee:2022nhh, Duhr:2022yyp, Moult:2022xzt}. 

The TMD distributions are specified by the TMD-twist, which is given by a pair of numbers $(n,m)$. In this nomenclature, ordinary TMD distributions of twist-two have TMD-twist-(1,1), which is specified in the index of TMD distributions in eq.~(\ref{WLP}). These TMD distributions are defined as
\begin{eqnarray}\label{def:PHI11}
\Phi_{11}^{[\Gamma]}(x,b)=\int_{-\infty}^\infty \frac{d\lambda}{2\pi} e^{-ix \lambda P^+} 
\langle P,s|\bar qW^\dagger(\lambda n+b)\frac{\Gamma}{2} Wq(0)|P,s\rangle,
\end{eqnarray}
where $|P,s\rangle$ is a hadron state with large component of momentum $P^+$, and
\begin{eqnarray}
W(x)=[-\infty n+x,x]=P\exp\(-ig \int_{-\infty}^0 d\sigma A_+(x+\sigma n)\)
\end{eqnarray}
is a straight Wilson line. The anti-quark distribution is defined as
\begin{eqnarray}\label{def:PHI11-anti}
\overline{\Phi}_{11}^{[\Gamma]}(x,b)&=&\int_{-\infty}^\infty \frac{d\lambda}{2\pi} e^{-ix \lambda P^+} 
\Tr\langle P,s|\frac{\Gamma}{2}Wq(\lambda n+b) \bar qW^\dagger(0)|P,s\rangle.
\end{eqnarray}
In both definitions we omit the renormalization constants and scaling arguments for simplicity. In what follows, we also consider the gluon TMD distribution of  twist-two. It is defined as
\begin{eqnarray}\label{def:PHI11-gluon}
\Phi_{11,\mu\nu}(x,b)=\frac{1}{xp_+}\int_{-\infty}^\infty \frac{d\lambda}{2\pi} e^{-ix \lambda P^+} 
\langle P,s|F_{\mu+}W^\dagger(\lambda n+b)\frac{\Gamma}{2} WF_{\nu+}(0)|P,s\rangle,
\end{eqnarray}
where $F_{\mu\nu}$ is the gluon field-strength tensor, and Wilson lines are in the adjoint color representation.

\subsubsection{Kinematic power corrections at NLP}
\label{sec:KPC1}

The KPC term can be found in ref.~\cite{Vladimirov:2021hdn}. It reads
\begin{eqnarray}\label{WKPC}
&&\widetilde{W}^{\mu\nu}_{\text{KPC}}
=\frac{i}{4}\mathbb{C}_{\text{LP}}\(\frac{\tau^2}{\mu^2}\)\sum_{n,m}\Bigg\{
\\\nn &&\phantom{+}
\frac{\bar n^\mu \Tr[\omega_{G}^{\rho}\overline\Gamma_m^- \omega_{G'}^\nu \overline\Gamma_n^+]+\bar n^\nu \Tr[\omega_{G}^\mu \overline\Gamma_m^-\omega_{G'}^\rho \overline\Gamma_n^+]}{q_-}
\Phi_{11}^{[\Gamma^+_n]}
\(\partial_\rho+\frac{\partial_\rho \mathcal{D}}{2}\ln\(\frac{\bar \zeta}{\zeta}\)\)
\overline{\Phi}_{11}^{[\Gamma_m^-]}
\\\nn &&
+\frac{\bar n^\mu \Tr[\omega_{G}^{\rho}\overline\Gamma_n^+ \omega_{G'}^\nu\overline\Gamma^-_m]
+
\bar n^\nu \Tr[\omega_{G}^\mu \overline\Gamma_n^+\omega_{G'}^\rho \overline\Gamma_m^-]}{q_-}
\overline{\Phi}_{11}^{[\Gamma_n^+]}
\(\partial_\rho+\frac{\partial_\rho \mathcal{D}}{2}\ln\(\frac{\bar \zeta}{\zeta}\)\)\Phi_{11}^{[\Gamma_m^-]}
\\\nn &&
+
\frac{n^\mu \Tr[\omega_{G}^{\rho}\overline\Gamma_m^- \omega_{G'}^\nu\overline\Gamma^+_n]
+
n^\nu \Tr[\omega_{G}^\mu \overline\Gamma_m^-\omega_{G'}^\rho \overline\Gamma_n^+]}{q_+}
\overline{\Phi}_{11}^{[\Gamma_m^-]}
\(\partial_\rho+\frac{\partial_\rho \mathcal{D}}{2}\ln\(\frac{\zeta}{\bar \zeta}\)\)\Phi_{11}^{[\Gamma_n^+]}
\\\nn &&
+
\frac{n^\mu \Tr[\omega_{G}^{\rho}\overline\Gamma_n^+ \omega_{G'}^\nu \overline\Gamma_m^-]
+
n^\nu \Tr[\omega_{G}^\mu \overline\Gamma_n^+\omega_{G'}^\rho \overline\Gamma_m^-]}{q_+}
\Phi_{11}^{[\Gamma^-_m]}
\(\partial_\rho+\frac{\partial_\rho \mathcal{D}}{2}\ln\(\frac{\zeta}{\bar \zeta}\)\)
\overline{\Phi}_{11}^{[\Gamma_n^+]}
\Bigg\}
,
\end{eqnarray}
where for shortness we have omitted the arguments of the TMD distributions, and $\partial_\rho=\partial/\partial b^\rho$. As so, all $\Phi^{[\Gamma^+_n]}$ and $\overline{\Phi}^{[\Gamma^+_n]}$ have argument $(x_1,b,\mu,\zeta)$, while all $\Phi^{[\Gamma^-_m]}$ and $\overline{\Phi}^{[\Gamma^-_m]}$ have argument $(x_2,b,\mu,\bar \zeta)$, which also defines the associated hadron. The function $\mathcal{D}(b,\mu)$ is the Collins-Soper kernel that dictates the evolution of TMD distributions with respect to the scale $\zeta$.

The defining feature of KPCs is that they involve only twist-two TMD distributions and their derivatives. That is, the KPC term does not introduce any new non-perturbative information. The hard coefficient function for the KPC term remains identical to that of the LP term, eq.~(\ref{WLP}), at all orders in perturbation theory. This follows from the requirement of electromagnetic gauge invariance, which dictates that the hadron tensor must be transverse to the vector $q^\mu$
$$
q_\mu W^{\mu\nu}=0.
$$
At LP, the direct computation yields
$$
q_\mu W_{\text{LP}}^{\mu\nu}\sim \frac{M}{Q}\neq 0.
$$
that is, there is a violation of  NLP order. At NLP it can be fixed by the KPC part of the hadronic tensor, because this is the only one that incorporates the same non-perturbative functions. Indeed, the direct computation shows
\begin{eqnarray}
q_\mu \(W_{\text{LP}}^{\mu\nu}+W^{\mu\nu}_{\text{KPC}}\)\sim \frac{M^2}{Q^2}\neq 0\,,
\end{eqnarray}
i.e. the electro-magnetic Ward identities violation is moved to  NNLP order. This chain continues to all powers of  TMD factorization theorem, as it has been demonstrated explicitly in ref.~\cite{Vladimirov:2023aot} and they must hold at all orders of perturbation theory. Therefore, the coefficient function for these terms is the same at all orders. In some sense, the KPCs can be considered as a descendant part of the LP term of the factorization theorem.

The combination of derivatives $\partial_\rho$ with derivatives of the Collins-Soper kernel is a consequence of the recombination of the so-called special rapidity divergences~\cite{Rodini:2022wic}. These terms are responsible for the restoration of  boost-invariance, which demands that the factorized result is independent of $\zeta$ and $\bar \zeta$ as long as the relation eq.~(\ref{zeta*zeta=tau2}) is satisfied. This implies the invariance under the rescaling $\zeta\to \alpha \zeta$ and $\bar \zeta\to \bar \zeta/\alpha$ with $\alpha>0$, which is satisfied by eq.~(\ref{WKPC}).

The ``long derivative'' is commuting with the evolution of TMD distributions. It is straightforward to see that
\begin{eqnarray}
\(\partial_\rho+\frac{\partial_\rho \mathcal{D}}{2}\ln\(\frac{\zeta}{\bar \zeta}\)\)\Phi(x,b;\mu,\zeta)
=
\(\partial_\rho+\frac{\partial_\rho \mathcal{D}}{2}\ln\(\frac{\zeta^2}{\tau^2}\)\)\Phi(x,b;\mu,\zeta),
\end{eqnarray}
where we used eq.~(\ref{zeta*zeta=tau2}). Evolving the distribution to an another scale $\zeta_1$ we get
\begin{eqnarray}
&&\(\partial_\rho+\frac{\partial_\rho \mathcal{D}}{2}\ln\(\frac{\zeta^2}{\tau^2}\)\)\(\frac{\zeta}{\zeta_1}\)^{-\mathcal{D}(b,\mu)}\Phi(x,b;\mu,\zeta_1)
\\\nn &&\qquad\qquad\qquad=
\(\frac{\zeta}{\zeta_1}\)^{-\mathcal{D}(b,\mu)}\(\partial_\rho+\frac{\partial_\rho \mathcal{D}}{2}\ln\(\frac{\zeta_1^2}{\tau^2}\)\)\Phi(x,b;\mu,\zeta_1).
\end{eqnarray}
In other words, the ``long derivative'' acting to the TMD distribution evolves as the TMD distribution itself, preserving the relation in eq.~(\ref{zeta*zeta=tau2}). This property is vital to guarantee the  factorization theorem.

\subsubsection{Genuine NLP part}

The genuine NLP part involves both twist-three and twist-two distributions. This contribution was originally derived in the operator form in ref.~\cite{Vladimirov:2021hdn}. Here, we present the genuine NLP part in a refined form, expressed in a more compact and explicit manner. It reads
\begin{eqnarray}\label{W_gNLP}
&&W_{\text{gNLP}}^{\mu\nu}=\frac{-i}{4}
\int_{-1}^1 du_1 d u_2 d u_3\delta(u_1+u_2+u_3)
\Bigg\{
\\\nn &&
\phantom{+}
\Big[\mathbb{C}_R T_-^{\mu\nu\rho}(\bar n,n)-i\pi \mathbb{C}_I T_+^{\mu\nu\rho}(\bar n,n)\Big]
\(\delta(x_2-u_3) \Phi_{11}^{[\Gamma^+_n]}\overline{\mathbf{\Phi}}_{\rho,\oplus}^{[\Gamma_m^-]}
+\delta(x_1-u_3)\mathbf{\Phi}_{\rho,\oplus}^{[\Gamma^+_n]}\overline{\Phi}_{11}^{[\Gamma_m^-]}\)
\\\nn
&&
-
\Big[i\mathbb{C}_R T_+^{\mu\nu\rho}(\bar n,n)+\pi \mathbb{C}_I T_-^{\mu\nu\rho}(\bar n,n)\Big]
\(\delta(x_2-u_3)\Phi_{11}^{[\Gamma^+_n]}\overline{\mathbf{\Phi}}_{\rho,\ominus}^{[\Gamma_m^-]}
+
\delta(x_1-u_3)\mathbf{\Phi}_{\rho,\ominus}^{[\Gamma^+_n]}\overline{\Phi}_{11}^{[\Gamma_m^-]}\)
\\\nn
&&
+
\Big[\mathbb{C}_R T_-^{\mu\nu\rho}(n,\bar n)-i\pi \mathbb{C}_I T_+^{\mu\nu\rho}(n,\bar n)\Big]
\(
\delta(x_2-u_3) \overline{\Phi}_{11}^{[\Gamma_n^+]}
\mathbf{\Phi}_{\rho,\oplus}^{[\Gamma^-_m]}
+
\delta(x_1-u_3)\overline{\mathbf{\Phi}}_{\rho,\oplus}^{[\Gamma_n^+]}\Phi_{11}^{[\Gamma^-_m]}
\)
\\\nn
&&
-\Big[i\mathbb{C}_R T_+^{\mu\nu\rho}(n,\bar n)+\pi \mathbb{C}_I T_-^{\mu\nu\rho}(n,\bar n)\Big]
\(
\delta(x_2-u_3) \overline{\Phi}_{11}^{[\Gamma_n^+]}\mathbf{\Phi}_{\rho,\ominus}^{[\Gamma^-_m]}
+
\delta(x_1-u_3) \overline{\mathbf{\Phi}}_{\rho,\ominus}^{[\Gamma_n^+]}\Phi_{11}^{[\Gamma^-_m]}
\)
\Bigg\},
\end{eqnarray}
where $\mathbf{\Phi}_{\rho,\bullet}$ are TMD distributions of twist-three defined below in eq.~(\ref{def:PHI21} - \ref{def:PHI-minus}).  Throughout the text we use $\bullet$ to indicate any secondary index of a twist-three distribution. Alike earlier, the arguments of TMD distributions are omitted for brevity. The products of the form  $\mathbf{\Phi}_{\bullet}\Phi_{11}$ have arguments as $\mathbf{\Phi}_{\bullet}(u_1,u_2,u_3,b;\mu,\zeta)\Phi_{11}(x_2,b;\mu,\bar \zeta)$, while products of the form
$\Phi_{11}\mathbf{\Phi}_{\bullet}$ have arguments as $\Phi_{11}(x_1,b;\mu,\zeta)\mathbf{\Phi}_{\bullet}(u_1,u_2,u_3,b;\mu,\bar \zeta)$. The tensors $T$ are defined as
\begin{eqnarray}
T_\pm^{\mu\nu\rho}(\bar n,n)&=&
\(\frac{\bar n^\mu }{q_-}-\frac{n^\mu}{q^+}\)\Tr[\omega_{G}^{\rho}\overline{\Gamma}_m^- \omega_{G'}^\nu \overline{\Gamma}_n^+]
\pm
\(\frac{\bar n^\nu}{q_-}-\frac{n^\nu}{q_+}\)\Tr[\omega_{G}^\mu \overline{\Gamma}_m^-\omega_{G'}^\rho \overline{\Gamma}_n^+],
\\
T_\pm^{\mu\nu\rho}(n,\bar n)&=&
\(\frac{n^\mu}{q_+}-\frac{\bar n^\mu}{q^-}\)\Tr[\omega_{G}^{\rho}\overline{\Gamma}_n^+ \omega_{G'}^\nu \overline{\Gamma}_m^-]
\pm
\(
\frac{n^\nu}{q_+}-\frac{\bar n^\nu}{q_-}\)\Tr[\omega_{G}^\mu \overline{\Gamma}_n^+\omega_{G'}^\rho \overline{\Gamma}_m^-].
\end{eqnarray}
The coefficient functions $\mathbb{C}_{R}$ and $\mathbb{C}_{I}$ originate from a single complex-valued coefficient function, derived at NLO in ref.~\cite{Vladimirov:2021hdn}. Functions $\mathbb{C}_{R}$ and $\mathbb{C}_{I}$ represent its real and imaginary parts, correspondingly. The details of the derivation can be found in appendix A of ref.~\cite{Rodini:2023plb}. Note that the derivation in ref.~\cite{Rodini:2023plb} is performed for the SIDIS case. In the DY case, the invariant mass $q^2$ is positive, which should be accounted in computation of the imaginary part of logarithms. We have obtained the following NLO expressions for the coefficient functions $\mathbb{C}_{R}$ and $\mathbb{C}_{I}$ in the DY process (this is one of the results of this paper),
\begin{eqnarray}
&&\mathbb{C}_{R}\(u_2,x,\frac{\tau^2}{\mu^2}\)=
\frac{1}{(u_2)_+}+a_s\Bigg\{
2\frac{C_F}{(u_2)_+}\(-\mathbf{L}^2+2\mathbf{L}-\frac{11}{2}+\frac{7\pi^2}{6}\)
\\\nn &&\qquad
+2\(C_F-\frac{C_A}{2}\)\frac{1}{(u_2)_+}\frac{x}{u_2}\[\(\mathbf{L}-2+\frac{1}{2}\ln\(\frac{|x+u_2|}{x}\)\)\ln\(\frac{|x+u_2|}{x}\)
+\frac{\pi^2}{2}\theta(-x-u_2)\]
\\\nn &&\qquad
+C_A\frac{x}{x+u_2}\[-\(\frac{\ln|u_2|}{u_2}\)_++\frac{\ln x}{(u_2)_+}
-\frac{\pi^2}{2}\delta(u_2)\]
\Bigg\}+\mathcal{O}(a_s^2)
\\
&&\mathbb{C}_{I}\(u_2,x,\frac{\tau^2}{\mu^2}\)=
\delta(u_2)+a_s\Bigg\{
2C_F\[\delta(u_2)\(-\mathbf{L}^2+2\mathbf{L}-\frac{15}{2}+\frac{7\pi^2}{6}\)+\frac{1}{(u_2)_+}\]
\\\nn &&\qquad
+2\(C_F-\frac{C_A}{2}\)\[\delta(u_2)\mathbf{L}
+\frac{1}{(u_2)_+}\frac{x}{u_2}\(
\theta(-x-u_2)(\mathbf{L}-2)-\theta(x+u_2)\ln\(\frac{|x+u_2|}{x}\)\)\]
\\\nn &&\qquad
+C_A\[\delta(u_2)(\ln x+2)-\frac{\theta(u_2)}{(u_2)_+}\frac{x}{x+u_2}\]\Bigg\}
+\mathcal{O}(a_s^2),
\end{eqnarray}
with $C_F=(N^2_c-1)/(2N_c)$, $C_A=N_c$ and $\mathbf{L}=\ln(\tau^2/\mu^2)$. The ``plus'' distribution is defined as
\begin{eqnarray}\label{plus-distr}
\int du \frac{f(u)}{(u)_+}=\int du \frac{f(u)-f(0)}{u}.
\end{eqnarray}
 The $\mu$-dependence cancels between the TMD distributions and the coefficient functions $\mathbb{C}_{R,I}$. An explicit demonstration of this cancellation at NLO can be found in refs.~\cite{Vladimirov:2021hdn, Rodini:2022wki}.

The twist-three TMDPDFs can have TMD-twist-(1,2) or (2,1). Correspondingly, they are defined as
\begin{eqnarray}\label{def:PHI21}
\Phi_{\mu,21}^{[\Gamma]}(x_1,x_2,x_3,b)&=&
\int_{-\infty}^\infty \frac{d\lambda_1 d\lambda_2}{(2\pi)^2} e^{i(x_1\lambda_1+x_2\lambda_2)P^+} 
\\\nn &&
\langle P,s|\bar q[\lambda_1n+b,\lambda_2n+b]F_{\mu+}W^\dagger (\lambda_2n+b)\frac{\Gamma}{2}Wq(0)|P,s\rangle,
\\\label{def:PHI12}
\Phi_{\mu,12}^{[\Gamma]}(x_1,x_2,x_3,b)&=&\int_{-\infty}^\infty \frac{d\lambda_1 d\lambda_2}{(2\pi)^2} e^{i(x_1\lambda_1+x_2\lambda_2)P^+} 
\\\nn &&
\langle P,s|\bar q W^\dagger(\lambda_1n+b)\frac{\Gamma}{2}W(\lambda_2n)F_{\mu+}[\lambda_2n,0]q(0)|P,s\rangle.
\end{eqnarray}
The anti-quark distributions are
\begin{eqnarray}
\label{def:PHI21-bar}
\overline{\Phi}_{\mu,21}^{[\Gamma]}(x_1,x_2,x_3,b)&=&\int_{-\infty}^\infty \frac{d\lambda_1 d\lambda_2}{(2\pi)^2} e^{i(x_1\lambda_1+x_2\lambda_2)P^+} 
\\\nn &&
\Tr\langle P,s|\frac{\Gamma}{2} W(\lambda_2 n+b) F_{\mu+}[\lambda_2 n+b,\lambda_1 n+b]q(\lambda_1 n+b)\bar qW^\dagger(0) |P,s\rangle,
\\\label{def:PHI12-bar}
\overline{\Phi}_{\mu,12}^{[\Gamma]}(x_1,x_2,x_3,b)&=&\int_{-\infty}^\infty \frac{d\lambda_1 d\lambda_2}{(2\pi)^2} e^{i(x_1\lambda_1+x_2\lambda_2)P^+} 
\\\nn &&
\Tr\langle P,s|\frac{\Gamma}{2} Wq(\lambda_1n+b)\bar q[0,\lambda_2n]F_{\mu+}W^\dagger(\lambda_2 n)|P,s\rangle,
\end{eqnarray}
where ``$\Tr$'' contracts spinor and color indices. In both cases, the variables $x_{1,2,3}$ are related by the equation $x_1+x_2+x_3=0$. Thus, the three-variable notation is redundant, but it is convenient to keep track of the symmetries of twist-three distributions. One can think at  $x_1$, $x_2$ and $x_3$  as the collinear momentum-fractions of partons, being $x_i$ positive for particles and  negative for anti-particles.
The parton interpretation and properties of twist-three distributions (including their evolution equations) are studied and presented in refs.~\cite{Rodini:2022wic, Rodini:2023plb}.

The TMD distributions of definite TMD-twist have autonomous evolution, in the sense that distributions $\Phi_{12}$ do not mix with distributions $\Phi_{21}$ during the change of scales $\mu$ or $\zeta$. However, these distributions are inconvenient to operate with,  because they are complex-valued functions with indefinite parity. It is more practical to introduce the T-parity-definite combinations
\begin{eqnarray}\label{def:PHI-plus}
\Phi_{\mu,\oplus}^{[\Gamma]}(x_{1,2,3};b)&=&\frac{\Phi_{\mu,21}^{[\Gamma]}(x_{1,2,3};b)+\Phi_{\mu,12}^{[\Gamma]}(-x_{3,2,1};b)}{2},
\\\label{def:PHI-minus}
\Phi_{\mu,\ominus}^{[\Gamma]}(x_{1,2,3};b)&=&i\frac{\Phi_{\mu,21}^{[\Gamma]}(x_{1,2,3};b)-\Phi_{\mu,12}^{[\Gamma]}(-x_{3,2,1};b)}{2},
\end{eqnarray}
and the same combinations for $\overline{\Phi}$. Here and in the following, we are using the shorthand notation for the series of arguments $x_{1,2,3}=(x_1,x_2,x_3)$ and correspondingly $-x_{3,2,1}=(-x_3,-x_2,-x_1)$. Under discrete transformations, $\Phi_{\oplus}$ and $\Phi_{\ominus}$ transform into themselves, making their interpretation somewhat clearer. Specifically, the $\Phi_{\oplus (\ominus)}(x_1,x_2,x_3)$ with $x_3>0$ represents the real (imaginary) part of the interference between a quark with collinear momentum fraction $x_3$ and a quark-gluon pair with momentum fractions $x_1$ and $x_2$ \cite{Rodini:2023plb}. A similar interpretation applies to the corresponding antiquark distributions.

Expressing the physical process cross-section in terms of $\Phi_{\oplus}$ and $\Phi_{\ominus}$ results in a natural structure. Each term of eq.~(\ref{W_gNLP}) is real-valued and involves a convolution of quark and antiquark distributions with positive $x$ and $u_3$. Consequently, each term can be directly associated with a specific partonic subprocess. However, a drawback of this construction is that $\Phi_{\oplus}$ and $\Phi_{\ominus}$ do not have a definite TMD-twist, meaning they do not evolve independently. Their evolution equations take a matrix form, leading to mixing between $\Phi_{\oplus}$ and $\Phi_{\ominus}$. Nevertheless, twist-three and twist-two distributions remain independent and do not mix with  each other.

One more detail is to be added to the definition of twist-three distributions. The bold-font distributions used in eq.~(\ref{W_gNLP}) differ from those introduced in eq.~(\ref{def:PHI21}-\ref{def:PHI12-bar}) by a subtraction term. The reason is that the terms of the factorization theorem are individually divergent at $u_2\to 0$ (the special rapidity divergence). This divergence is canceled in the sum of all terms, and thus does not violate the factorization theorem. Nonetheless, it prevents a definition of individual distributions. To eliminate the special rapidity divergences, and to make each term finite, one should explicitly subtract the divergences. This leads to the physical TMD distributions of twist-three, which are denoted by the bold font. Following~\cite{Rodini:2022wki}, we define 
\begin{eqnarray}\label{def:subtraction}
\mathbf{\Phi}_{\mu,\bullet}^{[\Gamma]}(x_{1,2,3},b)=
\Phi_{\mu,\bullet}^{[\Gamma]}(x_{1,2,3},b)-\lim_{|x_2|\to 0}[\mathcal{R}_{\bullet}\otimes \Phi_{11}]^{[\Gamma]}_\mu(x_{1,2,3},b),
\end{eqnarray}
where $\mathcal{R}$ are kernels that could be computed perturbatively, and $\otimes$ is an integral convolution. The kernels $\mathcal{R}$ are proportional to the derivative of the Collins-Soper kernel. At LO they read \cite{Rodini:2022wki}
\begin{eqnarray}\label{subtraction-term1}
[\mathcal{R}_{21}\otimes \Phi_{11}]_\mu^{[\Gamma]}(x_{1,2,3},b)&=&\partial_\mu \mathcal{D}(b)\Phi_{11}^{[\Gamma]}(-x_1,b)\(\theta(x_2,x_3)-\theta(-x_2,-x_3)\),
\\\,\label{subtraction-term2}
[\mathcal{R}_{12}\otimes \Phi_{11}]_\mu^{[\Gamma]}(x_{1,2,3},b)&=&\partial_\mu \mathcal{D}(b)\Phi_{11}^{[\Gamma]}(x_3,b)\(\theta(x_1,x_2)-\theta(-x_1,-x_2)\),
\end{eqnarray}
where $\theta(x_1,x_2)$ is the product of Heaviside functions $\theta(x_1)\theta(x_2)$. The limit $|x_2|\to0$  should be taken for each term independently, i.e. for the terms with $\theta(x_2)$ it is the limit $x_2\to +0$, while for the terms with $\theta(-x_2)$ it is the limit $x_2\to-0$. As a result of the subtraction, the distributions $\mathbf{\Phi}_{\mu,\bullet}$ becomes continuous at $x_2=0$ line, and thus the integration with $1/(u_2)_+$ in eq.~(\ref{W_gNLP}) is well-defined. The subtraction procedure, and the recombination of divergent $\mathcal{R}\otimes \Phi$-terms form contributions proportional to $\ln(\zeta/\bar \zeta)$, which, in turn, combine with ordinary derivatives within the KPC part and compose the ``long derivatives'' in eq.~(\ref{WKPC}). As we discussed in the previous section, this is crucial for TMD factorization at NLP, and it is responsible for the restoration the boost invariance and preservation of the evolution properties.

\subsection{Angular structure functions}
\label{sec:AS}

Given the hadron tensors, we can now compute the individual angular structure functions. It is instructive to present the contributions to these functions separately for the LP, KPC, and genuine NLP terms in this subsection.

Note that contracting the lepton and hadron tensors provides a result that is not homogeneous in powers of $Q$. This arises because the lepton tensor is ``exact", meaning it retains all powers of $Q$, whereas the hadron tensor is ordered in power counting. As a result of this direct contraction, several terms appear that are NNLP relative to the leading term. These terms are influenced by NNLP computations, and for this reason, they are usually discarded. However, for the sake of completeness, we present all terms, marking the induced NNLP contributions in color for clarity.

\subsubsection{Leading power term}

To compute the LP contribution we need the parametrization of twist-two TMD distributions. The standard parametrization reads 
\begin{eqnarray}\label{def:TMDs:1:g+}
\Phi^{[\gamma^+]}_{11}(x,b)&=&f_1(x,b)+...~,
\\\label{def:TMDs:1:g+5}
\Phi^{[\gamma^+\gamma^5]}_{11}(x,b)&=&...~,
\\\label{def:TMDs:1:s+}
\Phi^{[i\sigma^{\alpha+}\gamma^5]}_{11}(x,b)&=&i\epsilon_T^{\alpha\mu}b_\mu M h_1^\perp(x,b)+...~,
\end{eqnarray}
where dots denote the polarized terms that are not required for the present computation. The distributions $f_1$ and $h_1^\perp$ are the unpolarized and Boer-Mulders TMDPDF, respectively. All TMD distributions are dimensionsless real functions that depend on $\vec b^2$ (the argument $b$ is used for shortness). The factor $M$ has mass dimension, and could be associated with the mass of a hadron, but we keep it equal for all terms for simplicity.

Substituting eq.~(\ref{def:TMDs:1:g+}-\ref{def:TMDs:1:s+}) into eq.~(\ref{WLP}) and contracting with lepton tensor we obtain the angular structure functions (\ref{def:Sigma_n}). The non-zero angular coefficients are
\begin{eqnarray}\label{LP:U}
\Sigma^{\text{LP}}_U&=&\frac{8\pi \alpha_{\text{em}}^2}{3N_cs Q^2}\sum_{f,G,G'}Q^4\Delta_G^*\Delta_{G'}
\Big[
\\\nn &&
z_{+\ell}^{GG'}z_{+f}^{GG'}\(1+\colorThree{\frac{\vec q_T^2}{2Q^2}}\)\mathcal{J}_0[f_1f_1]
-
z_{+\ell}^{GG'}r_{+f}^{GG'}\colorThree{\frac{\vec q_T^2}{2Q^2}}\mathcal{J}_2[h^\perp_1h^\perp_1]\Big],
\\
\Sigma^{\text{LP}}_2&=&\frac{8\pi \alpha_{\text{em}}^2}{3N_cs Q^2}\sum_{f,G,G'}Q^4\Delta_G^*\Delta_{G'}
\Big[
\\\nn &&
-z_{+\ell}^{GG'}z_{+f}^{GG'}\colorThree{\frac{\vec q_T^2}{Q^2}}\mathcal{J}_0[f_1f_1]
+
z_{+\ell}^{GG'}r_{+f}^{GG'}\(2+\colorThree{\frac{\vec q_T^2}{Q^2}}\)\mathcal{J}_2[h^\perp_1h^\perp_1]\Big],
\\
\Sigma^{\text{LP}}_4&=&\frac{8\pi \alpha_{\text{em}}^2}{3N_cs Q^2}\sum_{f,G,G'}Q^4\Delta_G^*\Delta_{G'}
z_{-\ell}^{GG'}z_{-f}^{GG'}2\sqrt{1+\colorThree{\frac{\vec q_T^2}{Q^2}}}\mathcal{J}_0[\{f_1f_1\}_A],
\\\label{LP:5}
\Sigma^{\text{LP}}_5&=&\frac{8\pi \alpha_{\text{em}}^2}{3N_cs Q^2}\sum_{f,G,G'}Q^4\Delta_G^*\Delta_{G'}
iz_{+\ell}^{GG'}r_{-f}^{GG'}\sqrt{1+\colorThree{\frac{\vec q_T^2}{Q^2}}}\mathcal{J}_2[\{h^\perp_1h^\perp_1\}_A],
\end{eqnarray}
where $\vec q_T^2=-q^2_T>0$. The coupling constants and  the combinations of propagators are presented in appendix~\ref{app:couplings}. The convolution functionals $\mathcal{J}$ are defined as
\begin{eqnarray}\label{def:Jn-S}
\mathcal{J}_n[f_1f_1]&=&\mathbb{C}_{\text{LP}}\(\frac{\tau^2}{\mu^2}\)\int_0^\infty \frac{b\,db}{2\pi} (bM)^n J_n(b|\vec q_T|)
\\\nn &&\times
\(\frac{\tau^2}{\zeta_\mu}\)^{-2\mathcal{D}(b,\mu)}\Big[
f_{1}(x_1,b)\overline{f}_{1}(x_2,b)+\overline{f}_{1}(x_1,b)f_{1}(x_2,b)\Big],
\\\label{def:Jn-A}
\mathcal{J}_n[\{f_1f_1\}_A]&=&\mathbb{C}_{\text{LP}}\(\frac{\tau^2}{\mu^2}\)\int_0^\infty \frac{b\,db}{2\pi} (bM)^n J_n(b|\vec q_T|)
\\\nn &&\times
\(\frac{\tau^2}{\zeta_\mu}\)^{-2\mathcal{D}(b,\mu)}\Big[
f_{1}(x_1,b)\overline{f}_{1}(x_2,b)-\overline{f}_{1}(x_1,b)f_{1}(x_2,b)\Big],
\end{eqnarray}
where $J_n$ is the Bessel function. In eq.~(\ref{def:Jn-S}, \ref{def:Jn-A}), we have applied the $\zeta$-prescription, and extracted the TMD evolution factor explicitly. 

These formulas agree with previous computations up to the \colorThree{colored terms}, which are all proportional to $\vec{q}_T^2/Q^2$. These terms arise as predictions from the LP factorization theorem, due to the contraction of the strictly LP hadronic tensor with the all-power leptonic tensor. However, they represent only a subset of the full NNLP corrections and are influenced by the choice of the factorization frame (which, in this context, determines the definition of the vectors $n$ and $\bar{n}$; see~\cite{Manohar:2002fd, Marcantonini:2008qn, Vladimirov:2023aot} for discussions on frame dependence). Furthermore, some of these terms originate from the violation of electromagnetic gauge invariance. For instance, the double-Boer-Mulders term in eq.~(\ref{LP:U}) vanishes in a complete all-power computation of twist-two contributions~\cite{Piloneta:2024aac}. Consequently, these terms should not be considered as they stand, but in a more general contest.

Also let us note, that the structure function $\Sigma_5$ (\ref{LP:5}) is originated by the interference between $\gamma$ and $Z$ bosons (\ref{app:interf-term}). Due to it, it has an extra  kinematical suppression factor $\sim \Gamma_Z/M_Z$, in comparison to the other angular structure functions.

\subsubsection{Kinematic power correction}

To evaluate KPC part of the hadron tensor we should resolve the vector part of derivative of TMD distributions with respect to $b$.  Using the fact that distributions depend only on $b^2$, we find
\begin{eqnarray}
\label{eq:der}
\(\partial_\rho+\frac{\partial_\rho \mathcal{D}}{2}\ln\(\frac{\zeta}{\bar \zeta}\)\)\Phi_{11}^{[\gamma^+]}&=&
-b^\rho M^2 \mathring{f}_1+...~,
\\
\(\partial_\rho+\frac{\partial_\rho \mathcal{D}}{2}\ln\(\frac{\zeta}{\bar \zeta}\)\)\Phi_{11}^{[i\sigma^{\alpha+}\gamma^5]}&=&
i\epsilon^{\alpha\rho}_T M h_1^\perp - i \epsilon^{\alpha \mu}_T b_\mu b^\rho M^3 \mathring{h}^\perp_1+...~,
\end{eqnarray}
where  dots denote the terms that depend on hadron polarization, and we omit the arguments of distributions.  
The functions with a circle on top are defined as
\begin{eqnarray}
\mathring{F}(x,b)=\frac{2}{M^2}\(\frac{\partial}{\partial \vec b^2}+\frac{1}{2}\ln\(\frac{\zeta}{\bar \zeta}\)\(\frac{\partial \mathcal{D}(b)}{\partial \vec b^2}\)\)F(x,b),
\end{eqnarray}
and the minus sign that appears in front of them in eq.~(\ref{eq:der}) is due to $b^2=-\vec b^2$.

Computing the KPC contribution we obtain the following non-vanishing structure functions
\begin{eqnarray}
\Sigma^{\text{KPC}}_U&=&\frac{8\pi \alpha_{\text{em}}^2}{3N_cs Q^2}\sum_{f,G,G'}Q^4\Delta_G^*\Delta_{G'}\colorThree{\frac{\vec q_T^2}{Q^2}}
\Big[
-z_{+\ell}^{GG'}z_{+f}^{GG'}\mathcal{J}_0[f_1f_1]
+
z_{+\ell}^{GG'}r_{+f}^{GG'}\mathcal{J}_2[h^\perp_1h^\perp_1]\Big],
\\
\Sigma^{\text{KPC}}_1&=&\frac{8\pi \alpha_{\text{em}}^2}{3N_cs Q^2}\sum_{f,G,G'}Q^4\Delta_G^*\Delta_{G'} \frac{M}{Q}
\Big[
\\\nn &&
z_{+\ell}^{GG'}z_{+f}^{GG'}\mathcal{J}_1[f_1\mathring{f}_1-\mathring{f}_1f_1]
+
z_{+\ell}^{GG'}r_{+f}^{GG'}\mathcal{J}_1[\boldsymbol{b}^2M^2(h^\perp_1\mathring{h}^\perp_1-\mathring{h}^\perp_1h^\perp_1)]\Big],
\\
\Sigma^{\text{KPC}}_2&=&\frac{8\pi \alpha_{\text{em}}^2}{3N_cs Q^2}\sum_{f,G,G'}Q^4\Delta_G^*\Delta_{G'} \colorThree{\frac{2\vec q_T^2}{Q^2}}
\Big[
z_{+\ell}^{GG'}z_{+f}^{GG'}\mathcal{J}_0[f_1f_1]
-
z_{+\ell}^{GG'}r_{+f}^{GG'}\mathcal{J}_2[h^\perp_1h^\perp_1]\Big],
\\
\Sigma^{\text{KPC}}_3&=&\frac{8\pi \alpha_{\text{em}}^2}{3N_cs Q^2}\sum_{f,G,G'}Q^4\Delta_G^*\Delta_{G'} \frac{2M}{\sqrt{\tau^2}}
z_{-\ell}^{GG'}z_{-f}^{GG'}\mathcal{J}_1[\{f_1\mathring{f}_1-\mathring{f}_1f_1\}_A],
\\
\Sigma^{\text{KPC}}_4&=&\frac{8\pi \alpha_{\text{em}}^2}{3N_cs Q^2}\sum_{f,G,G'}Q^4\Delta_G^*\Delta_{G'} \colorThree{\frac{-2\vec q_T^2}{Q\sqrt{\tau^2}}}
z_{-\ell}^{GG'}z_{-f}^{GG'}\mathcal{J}_0[\{f_1f_1\}_A],
\\
\Sigma^{\text{KPC}}_5&=&\frac{8\pi \alpha_{\text{em}}^2}{3N_cs Q^2}\sum_{f,G,G'}Q^4\Delta_G^*\Delta_{G'} \colorThree{\frac{-\vec q_T^2}{Q\sqrt{\tau^2}}}
iz_{+\ell}^{GG'}r_{-f}^{GG'}\mathcal{J}_2[\{h^\perp_1 h^\perp_1\}_A],
\\
\Sigma^{\text{KPC}}_6&=&\frac{8\pi \alpha_{\text{em}}^2}{3N_cs Q^2}\sum_{f,G,G'}Q^4\Delta_G^*\Delta_{G'} \frac{M}{\sqrt{\tau^2}}
iz_{+\ell}^{GG'}r_{-f}^{GG'}\mathcal{J}_1[\boldsymbol{b}^2M^2\{h^\perp_1 \mathring{h}^\perp_1+\mathring{h}^\perp_1 h^\perp_1\}_A],
\end{eqnarray}
where we using the same notation as in eq.~(\ref{LP:U}-\ref{LP:5}). To compute these, we have integrated by parts and used the properties of the Bessel functions. Together they lead to the following relations
\begin{eqnarray}
\mathcal{J}_1[f_1\mathring{f}_2]&=&-\frac{|\vec q_T|}{M}\mathcal{J}_0[f_1f_2]-\mathcal{J}_1[\mathring{f}_1f_2],
\\
\mathcal{J}_1[\vec b^2 M^2 f_1\mathring{f}_2]&=&\frac{|\vec q_T|}{M}\mathcal{J}_2[f_1f_2]-4\mathcal{J}_1[f_1f_2]-\mathcal{J}_1[\vec b^2 M^2\mathring{f}_1f_2].
\end{eqnarray}
The integration by parts helps to refactor some NLP contributions, and to demonstrate their NNLP nature. The angular structures $\Sigma_{U,2,4,5}$ indicated \colorThree{in color} can be ignored, because they should combine with NNLP terms of LP and NNLP hadron tensors. Note, that these structure functions are exactly the same that have a non-zero NNLP component in eq.~(\ref{LP:U}-\ref{LP:5}). Meanwhile, the structure functions $\Sigma_{1,3,6}$ represent the true NLP contributions, and they do not have LP term.

\subsubsection{Genuine NLP part}

To process the genuine NLP part, we need a parameterization for the twist-three TMD distributions as introduced in ref.~\cite{Rodini:2022wki} and elaborated in ref.~\cite{Rodini:2023plb}. In total there are 32 quark TMDPDFs of twist-three. Here we use only the unpolarized ones,
\begin{eqnarray}\label{def:TMDs:2:g+}
\Phi_{\bullet}^{\mu[\gamma^+]}(x_{1,2,3},b)&=&
ib^\mu M^2 f^\perp_\bullet(x_{1,2,3},b)
+...~,
\\\label{def:TMDs:2:g+5}
\Phi_{\bullet}^{\mu[\gamma^+\gamma^5]}(x_{1,2,3},b)&=&
-i\epsilon^{\mu\nu}_Tb_\nu M^2 g^\perp_\bullet(x_{1,2,3},b)
+...~,
\\\label{def:TMDs:2:s+}
\Phi_{\bullet}^{\mu[i\sigma^{\alpha+}\gamma^5]}(x_{1,2,3},b)&=&
\epsilon^{\mu\alpha}_T M h_\bullet(x_{1,2,3},b)
+(b^\mu \epsilon^{\alpha\beta}_Tb_\beta+\epsilon_T^{\mu\beta}b_\beta b^\alpha)M^3h_{\bullet}^\perp(x_{1,2,3},b)
+...~
,
\end{eqnarray}
where $\bullet$ is $\oplus$ or $\ominus$, and dots denote the  terms that depend on hadron polarization. 

Inserting the parametrization of eq.~(\ref{def:TMDs:2:g+}-\ref{def:TMDs:2:s+}) into  the genuine NLP part, eq.~(\ref{W_gNLP}), we obtain the following non-zero angular coefficients
\begin{eqnarray}\label{gNLP:1}
\Sigma^{\text{gNLP}}_1&=&\frac{8\pi \alpha_{\text{em}}^2}{3N_cs Q^2}\sum_{f,G,G'}Q^4\Delta_G^*\Delta_{G'}
\frac{2M}{Q}\Big[
\\\nn && \phantom{+}
z_{+\ell}^{GG'}z_{+f}^{GG'}\mathcal{J}_1^R[\{f_1\mathbf{f}_\ominus^\perp+\mathbf{f}_\ominus^\perp f_1\}_A+\{f_1\mathbf{g}_\oplus^\perp-\mathbf{g}_\oplus^\perp f_1\}_S]
-
2z_{+\ell}^{GG'}r_{+f}^{GG'}\mathcal{J}_1^R[\{h_1^\perp\mathbf{h}_\ominus+\mathbf{h}_\ominus h_1^\perp\}_A]
\\\nn &&
+z_{+\ell}^{GG'}z_{+f}^{GG'}\mathcal{J}_1^I[\{f_1\mathbf{f}_\oplus^\perp+\mathbf{f}_\oplus^\perp f_1\}_A-\{f_1\mathbf{g}_\ominus^\perp-\mathbf{g}_\ominus^\perp f_1\}_S]
-
2z_{+\ell}^{GG'}r_{+f}^{GG'}\mathcal{J}_1^I[\{h_1^\perp\mathbf{h}_\oplus+\mathbf{h}_\oplus h_1^\perp\}_A]
\Big],
\\
\Sigma^{\text{gNLP}}_3&=&\frac{8\pi \alpha_{\text{em}}^2}{3N_cs Q^2}\sum_{f,G,G'}Q^4\Delta_G^*\Delta_{G'}
\frac{4M}{\sqrt{\tau^2}}\Big[
\\\nn &&
\phantom{+} 
z_{-\ell}^{GG'}z_{-f}^{GG'}\mathcal{J}_1^R[\{f_1\mathbf{f}_\ominus^\perp+\mathbf{f}_\ominus^\perp f_1\}_S+\{f_1\mathbf{g}_\oplus^\perp-\mathbf{g}_\oplus^\perp f_1\}_A]
+
2iz_{-\ell}^{GG'}r_{-f}^{GG'}\mathcal{J}_1^R[\{h_1^\perp\mathbf{h}_\oplus+\mathbf{h}_\oplus h_1^\perp\}_S]
\\\nn &&
+z_{-\ell}^{GG'}z_{-f}^{GG'}\mathcal{J}_1^I[\{f_1\mathbf{f}_\oplus^\perp+\mathbf{f}_\oplus^\perp f_1\}_S-\{f_1\mathbf{g}_\ominus^\perp-\mathbf{g}_\ominus^\perp f_1\}_A]
-
2iz_{-\ell}^{GG'}r_{-f}^{GG'}\mathcal{J}_1^I[\{h_1^\perp\mathbf{h}_\ominus+\mathbf{h}_\ominus h_1^\perp\}_S]
\Big],
\\
\Sigma^{\text{gNLP}}_6&=&\frac{8\pi \alpha_{\text{em}}^2}{3N_cs Q^2}\sum_{f,G,G'}Q^4\Delta_G^*\Delta_{G'}
\frac{2M}{\sqrt{\tau^2}}\Big[
\\\nn && \phantom{+}
z_{+\ell}^{GG'}z_{-f}^{GG'}\mathcal{J}_1^R[\{f_1\mathbf{f}_\oplus^\perp+\mathbf{f}_\oplus^\perp f_1\}_S-\{f_1\mathbf{g}_\ominus^\perp-\mathbf{g}_\ominus^\perp f_1\}_A]
-
2iz_{+\ell}^{GG'}r_{-f}^{GG'}\mathcal{J}_1^R[\{h_1^\perp\mathbf{h}_\ominus+\mathbf{h}_\ominus h_1^\perp\}_S]
\\\nn &&
-z_{+\ell}^{GG'}z_{-f}^{GG'}\mathcal{J}_1^I[\{f_1\mathbf{f}_\ominus^\perp+\mathbf{f}_\ominus^\perp f_1\}_S+\{f_1\mathbf{g}_\oplus^\perp-\mathbf{g}_\oplus^\perp f_1\}_A]
-
2iz_{+\ell}^{GG'}r_{-f}^{GG'}\mathcal{J}_1^I[\{h_1^\perp\mathbf{h}_\oplus+\mathbf{h}_\oplus h_1^\perp\}_S]
\Big],
\\\label{gNLP:7}
\Sigma^{\text{gNLP}}_7&=&\frac{8\pi \alpha_{\text{em}}^2}{3N_cs Q^2}\sum_{f,G,G'}Q^4\Delta_G^*\Delta_{G'}
\frac{4M}{Q}\Big[
\\\nn && 
\phantom{+}
z_{-\ell}^{GG'}z_{+f}^{GG'}\mathcal{J}_1^R[\{f_1\mathbf{f}_\oplus^\perp+\mathbf{f}_\oplus^\perp f_1\}_A-\{f_1\mathbf{g}_\ominus^\perp-\mathbf{g}_\ominus^\perp f_1\}_S]
-
2z_{-\ell}^{GG'}r_{+f}^{GG'}\mathcal{J}_1^R[\{h_1^\perp\mathbf{h}_\oplus+\mathbf{h}_\oplus h_1^\perp\}_A]
\\\nn &&
-z_{-\ell}^{GG'}z_{+f}^{GG'}\mathcal{J}_1^I[\{f_1\mathbf{f}_\ominus^\perp+\mathbf{f}_\ominus^\perp f_1\}_A+\{f_1\mathbf{g}_\oplus^\perp-\mathbf{g}_\oplus^\perp f_1\}_S]
+
2z_{-\ell}^{GG'}r_{+f}^{GG'}\mathcal{J}_1^I[\{h_1^\perp\mathbf{h}_\ominus+\mathbf{h}_\ominus h_1^\perp\}_A]
\Big].
\end{eqnarray}
The convolution integral are defined as
\begin{eqnarray}\label{def:J1-3}
\mathcal{J}_1^R[\{f_1f_\bullet\}_S]&=&\int_0^\infty \frac{b\,db}{2\pi} (bM) J_1(b|\vec q_T|)\(\frac{\tau^2}{\zeta_\mu}\)^{-2\mathcal{D}(b,\mu)}
\int_{-1}^{1-x_2} du_2 \mathbb{C}_{R}\(u_2,x_2;\frac{\tau^2}{\mu^2}\)
\Big[
\\\nn &&
f_{1}(x_1,b)\overline{f}_{\bullet}(-u_2-x_2,u_2,x_2,b)+\overline{f}_{1}(x_1,b)f_{\bullet}(-u_2-x_2,u_2,x_2,b)\Big],
\\\label{def:J1-33}
\mathcal{J}_1^R[\{f_\bullet f_1\}_S]&=&\int_0^\infty \frac{b\,db}{2\pi} (bM) J_1(b|\vec q_T|)\(\frac{\tau^2}{\zeta_\mu}\)^{-2\mathcal{D}(b,\mu)}
\int_{-1}^{1-x_1} du_2 \mathbb{C}_{R}\(u_2,x_1;\frac{\tau^2}{\mu^2}\)
\Big[
\\\nn &&
f_{\bullet}(-u_2-x_1,u_2,x_1,b)\overline{f}_{1}(x_2,b)+\overline{f}_{\bullet}(-u_2-x_1,u_2,x_1,b)f_{1}(x_2,b)\Big],
\end{eqnarray}
where $f_1$ is any TMD distribution of twist-two, and $f_\bullet$ is any distribution of twist-three. For convolutions $\mathcal{J}_1^I$ has the same form with the coefficient function $\mathbb{C}_R$ been replaced by $\pi \mathbb{C}_I$. For convolutions with label $\{..\}_A$ the quark and anti-quark distributions must be taken in the anti-symmetric combination, alike in the definition (\ref{def:Jn-A}).

The combinations of distributions that appear in eq.~(\ref{gNLP:1}-\ref{gNLP:7}) represent all possible linear independent combinations of $f^\perp_\bullet$, $g^\perp_\bullet$ and $h_\bullet$. They simplify by introducing the linear combinations
\begin{eqnarray}\label{combinations}
\mathbf{f}_{2}^\perp=\mathbf{f}_{\ominus}^\perp-\mathbf{g}_{\oplus}^\perp,\qquad \mathbf{g}_{2}^\perp=\mathbf{f}_{\oplus}^\perp+\mathbf{g}_{\ominus}^\perp,
\qquad
\overline{\mathbf{f}}_{2}^\perp=\overline{\mathbf{f}}_{\ominus}^\perp+\overline{\mathbf{g}}_{\oplus}^\perp,\qquad \overline{\mathbf{g}}_{2}^\perp=\overline{\mathbf{f}}_{\oplus}^\perp-\overline{\mathbf{g}}_{\ominus}^\perp,
\end{eqnarray}
where argument $(x_{1,2,3},b;\mu,\zeta)$ is omitted for all distributions. Note, that for the anti-quark distributions $\overline{\mathbf{f}}_{2}^\perp$ and $\overline{\mathbf{g}}_{2}^\perp$ the sign of the $g_\bullet$-part inverts. This is due to its Dirac structure $\gamma^+(1+\gamma^5)$, which turns to $\gamma^+(1-\gamma^5)$ under the conjugation. The combinations (\ref{combinations}) are related to the twist-three bi-quark TMD distributions $f^\perp$ and $g^\perp$  \cite{Mulders:1995dh, Bacchetta:2006tn, Rodini:2022wki}. 

In the terms of $\mathbf{f}_2^\perp$ and $\mathbf{g}_2^\perp$, eq.~(\ref{gNLP:1}-\ref{gNLP:7}) reads
\begin{eqnarray}\label{gNLP:1M}
\Sigma^{\text{gNLP}}_1&=&\frac{8\pi \alpha_{\text{em}}^2}{3N_cs Q^2}\sum_{f,G,G'}Q^4\Delta_G^*\Delta_{G'}
\frac{2M}{Q}\Big[
\\\nn && \phantom{+}
z_{+\ell}^{GG'}z_{+f}^{GG'}\mathcal{J}_1^R[\{f_1\mathbf{f}_2^\perp+\mathbf{f}_2^\perp f_1\}_A]
-
2z_{+\ell}^{GG'}r_{+f}^{GG'}\mathcal{J}_1^R[\{h_1^\perp\mathbf{h}_\ominus+\mathbf{h}_\ominus h_1^\perp\}_A]
\\\nn &&
+z_{+\ell}^{GG'}z_{+f}^{GG'}\mathcal{J}_1^I[\{f_1\mathbf{g}_2^\perp+\mathbf{g}_2^\perp f_1\}_A]
-
2z_{+\ell}^{GG'}r_{+f}^{GG'}\mathcal{J}_1^I[\{h_1^\perp\mathbf{h}_\oplus+\mathbf{h}_\oplus h_1^\perp\}_A]
\Big],
\\
\Sigma^{\text{gNLP}}_3&=&\frac{8\pi \alpha_{\text{em}}^2}{3N_cs Q^2}\sum_{f,G,G'}Q^4\Delta_G^*\Delta_{G'}
\frac{4M}{\sqrt{\tau^2}}\Big[
\\\nn && 
\phantom{+}
z_{-\ell}^{GG'}z_{-f}^{GG'}\mathcal{J}_1^R[\{f_1\mathbf{f}_2^\perp+\mathbf{f}_2^\perp f_1\}_S]
+
2iz_{-\ell}^{GG'}r_{-f}^{GG'}\mathcal{J}_1^R[\{h_1^\perp\mathbf{h}_\oplus+\mathbf{h}_\oplus h_1^\perp\}_S]
\\\nn &&
+z_{-\ell}^{GG'}z_{-f}^{GG'}\mathcal{J}_1^I[\{f_1\mathbf{g}_2^\perp+\mathbf{g}_2^\perp f_1\}_S]
-
2iz_{-\ell}^{GG'}r_{-f}^{GG'}\mathcal{J}_1^I[\{h_1^\perp\mathbf{h}_\ominus+\mathbf{h}_\ominus h_1^\perp\}_S]
\Big],
\\
\Sigma^{\text{gNLP}}_6&=&\frac{8\pi \alpha_{\text{em}}^2}{3N_cs Q^2}\sum_{f,G,G'}Q^4\Delta_G^*\Delta_{G'}
\frac{2M}{\sqrt{\tau^2}}\Big[
\\\nn && \phantom{-}
z_{+\ell}^{GG'}z_{-f}^{GG'}\mathcal{J}_1^R[\{f_1\mathbf{g}_2^\perp+\mathbf{g}_2^\perp f_1\}_S]
-
2iz_{+\ell}^{GG'}r_{-f}^{GG'}\mathcal{J}_1^R[\{h_1^\perp\mathbf{h}_\ominus+\mathbf{h}_\ominus h_1^\perp\}_S]
\\\nn &&
-z_{+\ell}^{GG'}z_{-f}^{GG'}\mathcal{J}_1^I[\{f_1\mathbf{f}_2^\perp+\mathbf{f}_2^\perp f_1\}_S]
-
2iz_{+\ell}^{GG'}r_{-f}^{GG'}\mathcal{J}_1^I[\{h_1^\perp\mathbf{h}_\oplus+\mathbf{h}_\oplus h_1^\perp\}_S]
\Big],
\\\label{gNLP:7M}
\Sigma^{\text{gNLP}}_7&=&\frac{8\pi \alpha_{\text{em}}^2}{3N_cs Q^2}\sum_{f,G,G'}Q^4\Delta_G^*\Delta_{G'}
\frac{4M}{Q}\Big[
\\\nn && 
\phantom{+}
z_{-\ell}^{GG'}z_{+f}^{GG'}\mathcal{J}_1^R[\{f_1\mathbf{g}_2^\perp+\mathbf{g}_2^\perp f_1\}_A]
-
2z_{-\ell}^{GG'}r_{+f}^{GG'}\mathcal{J}_1^R[\{h_1^\perp\mathbf{h}_\oplus+\mathbf{h}_\oplus h_1^\perp\}_A]
\\\nn &&
-z_{-\ell}^{GG'}z_{+f}^{GG'}\mathcal{J}_1^I[\{f_1\mathbf{f}_2^\perp+\mathbf{f}_2^\perp f_1\}_A]
+
2z_{-\ell}^{GG'}r_{+f}^{GG'}\mathcal{J}_1^I[\{h_1^\perp\mathbf{h}_\ominus+\mathbf{h}_\ominus h_1^\perp\}_A]
\Big].
\end{eqnarray}
The pairs $(\mathbf{f}_2^\perp, \mathbf{g}_2^\perp)$ and $(\mathbf{h}_\oplus, \mathbf{h}_\ominus)$ are eigen-pairs of the twist-three evolution equation~\cite{Rodini:2022wki}. In other words, these distributions mix with each other within the pair, but do not mix with other distributions. This fact is also evident in the structure of the genuine terms of the factorization theorem, eq.~(\ref{gNLP:1M}-\ref{gNLP:7M}), where corresponding terms enter symmetrically into $\mathcal{J}^R$ and $\mathcal{J}^I$.

The twist-three functions $\mathbf{h}_\oplus^\perp$, $\mathbf{h}_\ominus^\perp$, as well as the combinations $\mathbf{f}_\oplus^\perp-\mathbf{g}^\perp_\ominus$ and $\mathbf{f}_\ominus^\perp+\mathbf{g}^\perp_\oplus$ do not contribute to the factorization at NLP with a spin-1 current, because they require a probe of spin-2 in order to have a non-zero projection (see discussion in ref.~\cite{Rodini:2022wki}). 

\section{Leading TMD approximation}
\label{sec:L-TMD}

The NLP formulas presented in section~\ref{sec:AS} are complete and ready for application. This is the first time they are formulated with a clear separation of definite TMD-twist distributions and with NLO coefficient functions. However, these formulas are not of immediate practical usage, since they involve four unknown non-perturbative functions: $\mathbf{f}_2^\perp$, $\mathbf{g}_2^\perp$, $\mathbf{h}_\oplus^\perp$, and $\mathbf{h}_\ominus^\perp$, in addition to the familiar unpolarized $f_1$ and Boer-Mulders $h_1^\perp$ distributions.  

In principle, a combination of measurements of structure functions $A_{1,3,6,7}$ could provide some insight into these twist-three distributions. A further difficulty is to disentangle flavor contributions and collinear momentum fractions, that could be resolved only by adding more observables of different kinds. To overcome this situation, and avoid the extraction the twist-three distributions, one should introduce a reasonable approximation for these distributions.  

A key to solve the problem comes from the observation that twist-three distributions exhibit a singular behavior at $b \to 0$. This singularity is implicit and it can be revealed only in a dedicated analysis of the small-$b$ asymptotic. The presence of this singularity significantly alters the behavior of power corrections. Instead of the naive expectation of a $\Lambda/Q$ suppression, the actual correction scales as $q_T/Q$, making it significant for phenomenology. Therefore, it is crucial to separate out this singular part from the rest of the contribution.  

In this section, we propose an approximation that extracts the most singular part of twist-three TMD distributions and parametrizes it in terms of twist-two distributions. In this way, the proposed approximation resolves both of the aforementioned problems. The suggested approximation is not unique since there is some freedom in its construction. Nonetheless, it is systematic, as it can be improved order-by-order in perturbation theory and  extended to higher-twist distributions. For brevity, we refer to it as the \textit{leading-TMD} (L-TMD) approximation, as it expresses the leading part of a TMD distribution in terms of other TMD distributions.

\subsection{General discussion}
\label{sec:L-TMD-general}

Twist-two TMD distributions have a smooth behavior in the limit $b^2\to0$. This limit can be systematically analyzed using the operator product expansion (OPE), which leads to 
\begin{eqnarray}\label{OPE-tw2}
\Phi_{11}(x,b;\mu,\zeta)=C_{\Phi/f}(\mathbf{L}_b;\mu,\zeta;\mu_{\text{OPE}})\otimes f_{\text{coll}}(x,\mu_{\text{OPE}})+\mathcal{O}(b^2),
\end{eqnarray}
where $f_{\text{coll}}$ is a collinear distribution, $C$ is the coefficient function and $\mathbf{L}_b=\ln(\mu_{\text{OPE}}^2 \vec b^2)$. The scale $\mu_{\text{OPE}}$ is the scale of OPE, whose dependence cancels between $C$ and $f$. The symbol $\otimes$ denotes the convolution over momentum fractions between the collinear distribution and the corresponding variables of the coefficient function. The type of collinear distribution $f_{\text{coll}}$ depends on the nature of the TMD distribution. For instance, in the case of the unpolarized TMD distribution $f_1$, the corresponding collinear counterpart is the unpolarized collinear parton distribution function (PDF) of twist two. For the Sivers function $f_{1T}^\perp$, the relevant collinear distributions are the twist-three distributions $T$ and $\Delta T$. A comprehensive list of such relations can be found in refs.~\cite{Moos:2020wvd, Boussarie:2023izj}. The OPE for twist-two TMD distributions, as given in eq.~(\ref{OPE-tw2}), has been extensively studied over the last decade and is known at least at NLO for all major distributions (see \cite{Kanazawa:2015ajw, Bacchetta:2013pqa, Dai:2014ala, Echevarria:2016scs, Gutierrez-Reyes:2018iod, Gutierrez-Reyes:2019rug, Scimemi:2019gge, Ebert:2020qef, Luo:2020epw, Ebert:2020yqt, Rein:2022odl}). The only exception is the pretzelosity function, which has been derived only at LO \cite{Moos:2020wvd}. These relations play a crucial role in the construction of phenomenological models for TMD distributions, making them highly valuable for practical applications.

The small-$b$ asymptotic of higher-twist TMD distributions differs from that of twist-two TMD distributions. Most of the 32 twist-three TMD distributions exhibit a smooth $b \to 0$ limit, similar to the twist-two case in eq.~(\ref{OPE-tw2}). However, some of them behave as  
\begin{eqnarray}\label{OPE-tw3}
\Phi_{\mu\bullet}(x_{1,2,3},b;\mu,\zeta)=\frac{b_\mu}{b^2}C_{\Phi_\bullet/f}(\mathbf{L}_b;\mu,\zeta;\mu_{\text{OPE}})\otimes f_{\text{coll}}(x,\mu_{\text{OPE}})+\mathcal{O}(b^0),
\end{eqnarray} 
where $\Phi_{\mu\bullet}$ represents a twist-three TMD distribution, and the rest of the notation follows that of eq.~(\ref{OPE-tw2}) with the caveat that in this case the collinear distribution $f_{\text{coll}}$ appearing in eq.~(\ref{OPE-tw3}) is necessarily of twist-two. This behavior arises because twist-three distributions are defined via three-point operators (see eq.~(\ref{def:PHI21}-\ref{def:PHI21-bar})), whereas twist-two distributions originate from two-point operators. The LO diagrams that describe such reduction are shown in fig.~\ref{fig:diag}, and by dimensional analysis, they scale as $1/b$. The terms proportional to $b^0$ in eq.~(\ref{OPE-tw3}) follow the standard OPE rules and can involve distributions of twist-two, twist-three, and twist-four.

As discussed above, the fact that twist-three distributions exhibit singular behavior in the small-$b$ limit is problematic and disrupts the natural power counting of corrections. To address this issue, we aim to separate the singular part and define a higher-twist distributions without this singular contribution. Explicitly, we introduce a new twist-three TMD distribution
\begin{eqnarray}\label{prop1}
\widehat{\Phi}_{\mu\bullet}(x_{1,2,3},b;\mu,\zeta)=\Phi_{\mu\bullet}(x_{1,2,3},b;\mu,\zeta)-\frac{b^\nu}{b^2}[S\otimes \Phi_{11}]_{\mu\nu}(x_{1,2,3},b;\mu,\zeta),
\end{eqnarray}
where $S\otimes \Phi_{11}$ is a convolution of some known expression with a twist-two TMD distribution and
$$\lim_{b^2\to0}\widehat{\Phi}_{\bullet}(x_{1,2,3},b;\mu,\zeta)=\text{finite}.$$
Note, that it is expected that all functions are well-defined in the full range of $b$. Therefore, the singularity $1/b$ is explicitly revealed, and can be treated integrating the hadron tensor over $b$. Such a decomposition can be made in various manners. Here we propose a scheme that looks the most suitable to our current perspective.

Each term of the decomposition in eq.~(\ref{prop1}) must respect all perturbative properties of the TMD distributions, otherwise such decomposition would violate the factorization theorem. The main property to respect is the TMD evolution. Each function of eq.~(\ref{prop1}) must obey the same evolution equations with respect to $\mu$ and $\zeta$, in order to keep $\widehat{\Phi}$ and $[S\otimes \Phi_{11}]$ independent at each scale. This is rather difficult to achieve practically, because the function $[S\otimes \Phi_{11}]$ incorporates non-perturbative parts of different flavors. Thus, one faces a problem of matching rapidity evolution for quarks and gluons distributions, which are given by different Collins-Soper kernels. To avoid this problem, we utilize the optimal definition\footnote{
One can also define the decomposition with generic scales. However, it requires the introduction of intermediate scales $(\mu,\zeta)$ at which the quark TMD distribution is matched onto the gluon TMD distribution. In the $\zeta$-prescription, this scale is the saddle-point of the field of anomalous dimensions, which is unique and fundamental.
} of the TMD distributions \cite{Scimemi:2018xaf}. Namely, before the redefinition of $\Phi_\bullet$, we evolve it to the scale $\zeta_\mu$, which is defined such that the double-logarithm part of the evolution vanishes and the TMD distributions obey ordinary renormalization group equations\footnote{
For the TMD distributions of twist-two there exists $\zeta_\mu$ such that the TMD distribution is exactly scaleless. For higher-twist distributions such curve is absent, because their anomalous dimensions depend on $x$. However, the double-logarithm part of the evolution equations, which mixes $\mu$ and $\zeta$ scaling, is $x$ independent~\cite{Rodini:2022wki}, and thus can be eliminated. Moreover, it is eliminated using the same $\zeta_\mu$ as for TMD distributions of twist-two.
}. Thus the problem simplifies to
\begin{eqnarray}\label{prop2}
\widehat{\Phi}_{\mu\bullet}(x_{1,2,3},b;\mu)=\Phi_{\mu\bullet}(x_{1,2,3},b;\mu)-\frac{b^\nu}{b^2}[S\otimes \Phi_{11}]_{\mu\nu}(x_{1,2,3},b;\mu),
\end{eqnarray}
where the parameter $\zeta$ is absent because of the definition of the $\zeta$-prescription.

In the small-$b$ regime, the OPE for the optimal twist-two distribution is
\begin{eqnarray}
\Phi_{11}(x,b)=C_{\Phi_{11}/f}(\mathbf{L}_b;\mu_{\text{OPE}})\otimes f_{\text{coll}}(x,\mu_{\text{OPE}})+\mathcal{O}(b^2),
\end{eqnarray}
where scales $\mu$ and $\zeta$ are absent due to the definition of the optimal TMD distribution, and the coefficient function does not have any double-logarithm. This series can be inverted perturbatively as
\begin{eqnarray}\label{OPE-invert}
f_{\text{coll}}(x,\mu)&=&C^{-1}_{\Phi_{11}/f}(\mathbf{L}_b;\mu)\otimes \Phi_{11}(x,b)+\mathcal{O}(b^2).
\end{eqnarray}
The corresponding OPE for optimal twist-three distribution is
\begin{eqnarray}\label{OPE-tw3-optimal}
\Phi_{\mu\bullet}(x_{1,2,3},b;\mu)=\frac{b_\nu}{b^2}C^{\mu\nu}_{\Phi_{\bullet}/f}(\mathbf{L}_b;\mu)\otimes f_{\text{coll}}(x,\mu)+\mathcal{O}(b^0),
\end{eqnarray}
where we have equalized scales $\mu$ and $\mu_{\text{OPE}}$. Then substituting formally eq.~(\ref{OPE-invert}) into eq.~(\ref{OPE-tw3-optimal}) we obtain a result with all the desired properties, i.e.
\begin{eqnarray}\label{def:S}
[S\otimes \Phi_{11}]^{\mu\nu}(x_{1,2,3},b;\mu)=C^{\mu\nu}_{\Phi_{\bullet}/f}(\mathbf{L}_b;\mu)\otimes C^{-1}_{\Phi_{11}/f}(\mathbf{L}_b;\mu)\otimes \Phi_{11}(x,b).
\end{eqnarray}
This $[S\otimes \Phi_{11}]$ has a well defined non-perturbative behavior in a full range of $b$; it is made from known functions; it  subtracts exactly the singular part of $\Phi_{\mu\bullet}$; and, importantly, the contracted $[S\otimes \Phi_{11}]_\mu$ obeys the correct evolution equation, which is ensured by the OPE. Finally, one can systematically improve the definition of $[S\otimes \Phi_{11}]_\mu$ computing the higher perturbative orders of the coefficient functions.

The drawback of the construction eq.~(\ref{def:S}) is that it is not unique. Indeed, the inversion formula eq.~(\ref{OPE-invert}) is ambiguous because one can modify the definition of $b$ on the right-hand-side hiding the redefinition parts into $\mathcal{O}(b^2)$. This ambiguity can be mitigated by  requesting extra properties of subtraction procedure, for example, demanding exact subtraction of $\sim b^2$ correction, or of target-mass corrections. Nonetheless, this problem is not severe, because $\widehat{\Phi}_{\mu\bullet}$ (which is influenced by such redefinition) contribute only at low $Q$.

Let us also mention that similar subtraction procedures can be formulated systematically for all higher twist TMD distributions. Generally,  higher-twist TMD distributions can have a higher order singularity at $b\to0$. These singularities can be subtracted in a similar fashion order-by-order in powers of $b$, generating finite TMD distributions. This procedure preserves all original non-perturbative information without any loss, although it redistributes it among different factorized parts.

\subsection{L-TMD approximation at LO}
\label{sec:L-TMD-LO}

\begin{figure}[t]
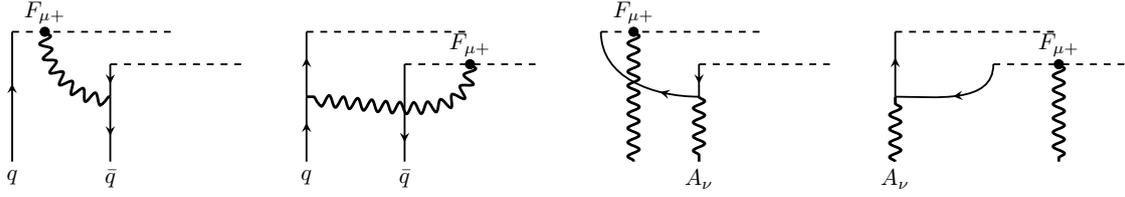

\begin{center}
\includestandalone[width=0.98\textwidth]{Figures/diagrams1}
\caption{\label{fig:diag} LO diagrams that produce the $\sim 1/b$ term in the small-$b$ OPE of TMD twist-three operators.}
\end{center}
\end{figure}

The diagrams that generate the LO singular term in the small-$b$ OPE are shown in fig.~\ref{fig:diag}. They present the matching of three-point $(\bar q Fq)$-operator to two-point $(\bar q q)$ and $(FF)$-operators. These diagrams are given in refs.~\cite{Rodini:2022wki, delCastillo:2023rng}. Since the quark and gluon-channel diagrams have different structures we discuss them one-by-one.

The diagrams with two quark fields are
\begin{eqnarray}\label{matching:12-sing}
\Phi^{[\Gamma]}_{\mu,12}(x_{1,2,3},b)\Big|_{\text{quark}}&=&a_s(\mu)\frac{C_F}{2}\frac{b^\nu}{b^2}\(\theta(x_1,x_2)-\theta(-x_1,-x_2)\)
\\\nn &&
\times
\sum_k
\Tr\[\overline{\Gamma}^+_k \(\gamma_\mu\gamma_\nu -\frac{x_1}{x_3}\gamma_\nu\gamma_\mu\)\Gamma\]
\phi^{[\Gamma^+_k]}(x_3,\mu)+\mathcal{O}\(b^0,\frac{a_s^2}{b}\),
\\\label{matching:21-sing}
\Phi^{[\Gamma]}_{\mu,21}(x_{1,2,3},b)\Big|_{\text{quark}}&=&a_s(\mu)\frac{C_F}{2}\frac{b^\nu}{b^2}\(\theta(x_2,x_3)-\theta(-x_2,-x_3)\)
\\\nn &&
\times
\sum_k
\Tr\[\Gamma\(\gamma_\nu\gamma_\mu -\frac{x_3}{x_1}\gamma_\mu\gamma_\nu\)\overline{\Gamma}^+_k\]
\phi^{[\Gamma^+_k]}(-x_1,\mu)+\mathcal{O}\(b^0,\frac{a_s^2}{b}\),
\end{eqnarray}
where $\Gamma\in\Gamma^+$. The collinear distribution is defined as
\begin{eqnarray}
\phi^{[\Gamma]}(x)&=&\int_{-\infty}^\infty \frac{d\lambda}{2\pi} e^{-ix \lambda P^+} 
\langle P,s|\bar q(\lambda n)\frac{\Gamma}{2}[\lambda n,0]q(0)|P,s\rangle.
\end{eqnarray}
The corrections to these expressions are either $\sim a_s^2/b$ (i.e. involving two-loop diagrams), or $\sim b^0$ (at the tree-order). 

To find the L-TMD approximation we must invert the collinear distribution in terms of TMD-distributions as in eq.~(\ref{OPE-invert}). For that we use the small-$b$ relation between twist-two TMD and collinear distributions. At LO it is
\begin{eqnarray}\label{matching:11}
\Phi^{[\Gamma]}_{11}(x,b)=\phi^{[\Gamma]}(x,\mu)+\mathcal{O}(a_s),
\end{eqnarray}
where $\Gamma\in\Gamma^+$. Consequently, the L-TMD approximation at LO for $\Phi_{12}$ and $\Phi_{21}$ is given simply by eq.~(\ref{matching:12-sing}-\ref{matching:21-sing}) with $\phi\to\Phi_{11}$. 

Both eq.~(\ref{matching:12-sing}) and (\ref{matching:21-sing}) have a discontinuity along the line $x_2=0$. This is the manifestation of the special rapidity divergence, as discussed in sec.~\ref{sec:KPC1}. To make the factorization theorem well-defined, one must apply the subtraction of eq.~(\ref{def:subtraction}). Using that
\begin{eqnarray}
\partial_\mu \mathcal{D}(b,\mu)=4a_s(\mu)C_F\frac{b_\mu}{b^2}+\mathcal{O}(a_s^2,b^0),
\end{eqnarray}
and rewriting the subtraction terms, eq.~(\ref{subtraction-term1}-\ref{subtraction-term2}) in the same form as in eq.~(\ref{matching:12-sing}- \ref{matching:21-sing}), the physical TMD distributions are found to be
\begin{eqnarray}\label{matching:12}
&&\mathbf{\Phi}^{[\Gamma]}_{\mu,12}(x_{1,2,3},b)\Big|_{\text{quark}}=a_s(\mu)\frac{C_F}{2}\frac{b^\nu}{b^2}\(\theta(x_1,x_2)-\theta(-x_1,-x_2)\)
\\\nn &&
\times
\sum_k\Big(
\Tr\[\overline{\Gamma}^+_k \(\gamma_\mu\gamma_\nu -\frac{x_1}{x_3}\gamma_\nu\gamma_\mu\)\Gamma\]
\Phi_{11}^{[\Gamma^+_k]}(x_3,b)
-2g_{\mu\nu}\Tr\[\overline{\Gamma}^+_k \Gamma\]
\Phi_{11}^{[\Gamma^+_k]}(-x_1,b)\Big)
+\mathcal{O}\(b^0,\frac{a_s^2}{b}\),
\\\label{matching:21}
&&\mathbf{\Phi}^{[\Gamma]}_{\mu,21}(x_{1,2,3},b)\Big|_{\text{quark}}=a_s(\mu)\frac{C_F}{2}\frac{b^\nu}{b^2}\(\theta(x_2,x_3)-\theta(-x_2,-x_3)\)
\\\nn &&
\times
\sum_k\Big(
\Tr\[\Gamma\(\gamma_\nu\gamma_\mu -\frac{x_3}{x_1}\gamma_\mu\gamma_\nu\)\overline{\Gamma}^+_k\]
\Phi_{11}^{[\Gamma^+_k]}(-x_1,b)
-2g_{\mu\nu}\Tr\[\overline{\Gamma}^+_k \Gamma\]
\Phi_{11}^{[\Gamma^+_k]}(x_3,b)\Big)
+\mathcal{O}\(b^0,\frac{a_s^2}{b}\),
\end{eqnarray}
where we used that $-x_1-x_3=x_2$, and consequently, $\lim_{|x_2|\to0}x_1=-x_3$ and $\lim_{|x_2|\to0}x_3=-x_1$. Due to the subtraction expressions (\ref{matching:12}, \ref{matching:21}) are zero at $x_2=0$. As the result, their contribution vanishes in the integral with $\delta(u_2)$, i.e. at the leading order of $\mathbb{C}_I$.

The matching to the gluon channel is the same for $\Phi_{\mu,12}$ and $\Phi_{\mu,21}$. It reads
\begin{eqnarray}\nn
\Phi^{[\Gamma]}_{\mu,12}(x_{1,2,3},b)\Big|_{\text{gluon}}&=&
\Phi^{[\Gamma]}_{\mu,21}(x_{1,2,3},b)\Big|_{\text{gluon}}
\\\label{matching:12-gluon}
&=&
-\frac{a_s(\mu)}{4}\frac{b_\nu}{b^2}\(\theta(x_1,x_3)-\theta(-x_1,-x_3)\)
\\\nn &&
\times
\Tr\[\(\frac{x_1}{x_2}\gamma^\nu\gamma^\alpha -\frac{x_3}{x_2}\gamma^\alpha\gamma^\nu\)\gamma^-\Gamma\]
\phi_{\mu\alpha}(-x_2,\mu)+\mathcal{O}\(b^0,\frac{a_s^2}{b}\),
\end{eqnarray}
where the collinear gluon distribution is
\begin{eqnarray}
\phi_{\mu\nu}(x)&=&\frac{1}{xp_+}\int_{-\infty}^\infty \frac{d\lambda}{2\pi} e^{-ix \lambda P^+} 
\langle P,s|F_{\mu+}(\lambda n)[\lambda n,0]F_{\nu+}(0)|P,s\rangle.
\end{eqnarray}
The gluon collinear distribution is related to the TMD distribution via the LO matching \cite{Echevarria:2015uaa, Gutierrez-Reyes:2019rug}
\begin{eqnarray}\label{matching:11-gluon}
\Phi_{11,\mu\nu}(x,b)=\phi_{\mu\nu}(x,\mu)+\mathcal{O}(a_s).
\end{eqnarray}
Thus, at LO, one can simply replace $\phi$ by $\Phi$, analogously to the quark case. The subtraction procedure does not affect the gluon channel.

We again emphasize that the direct matching between quark and gluon distributions can be made because of the $\zeta$-prescription. Due to it, both eq.~(\ref{matching:11-gluon}) and eq.~(\ref{matching:12}- \ref{matching:21}) are $\zeta$-independent. In the case of general scales, one must take into account that the $\zeta$-evolution of quark and gluon TMD distribution is different, and cannot be matched by perturbation theory (due to the non-perturbative nature of the Collins-Soper kernel). The matching can be done  introducing  some intermediate reference scale to which both distributions are evolved, with the price of having an extra undesired degree of freedom in the approximation. 

Combining eq.~(\ref{matching:11-gluon}) and (\ref{matching:12}-\ref{matching:21}) together we obtain the   twist-three TMD distributions in the L-TMD approximation. It is convenient to present them in the form
\begin{eqnarray}
\mathbf{\Phi}_{\mu,12}^{[\Gamma]}(x_{1,2,3},b)&=&
\widehat{\mathbf{\Phi}}_{\mu,12}^{[\Gamma]}(x_{1,2,3},b)
\\\nn&&
+\frac{b^\nu}{b^2}\int dy\(\sum_k  S_{\mu\nu}^{\Gamma,\overline{\Gamma}^+_k}(x_{1,2,3},y) \Phi_{11}^{[\Gamma_k^+]}(y,b)
+S_{\mu\nu,\alpha\beta}^{\Gamma}(x_{1,2,3},y) \Phi_{11}^{\alpha\beta}(y,b)\),
\\
\mathbf{\Phi}_{\mu,21}^{[\Gamma]}(x_{1,2,3},b)&=&
\widehat{\mathbf{\Phi}}_{\mu,21}^{[\Gamma]}(x_{1,2,3},b)
\\\nn&&
+\frac{b^\nu}{b^2}\int dy\(-\sum_k  S_{\mu\nu}^{\dagger\Gamma,\overline{\Gamma}^+_k}(-x_{3,2,1},y) \Phi_{11}^{[\Gamma_k^+]}(y,b)+S_{\mu\nu,\alpha\beta}^{\Gamma}(x_{1,2,3},y) \Phi_{11}^{\alpha\beta}(y,b)\),
\end{eqnarray}
where
\begin{eqnarray}\label{def:S1}
&&S_{\mu\nu}^{\Gamma,\overline{\Gamma}^+_k}(x_{1,2,3},y)=a_s(\mu)\frac{C_F}{2}\(\theta(x_1,x_2)-\theta(-x_1,-x_2)\)
\\\nn &&\qquad\qquad
\Big(
\Tr\[\overline{\Gamma}^+_k\(\gamma_\mu\gamma_\nu-\frac{x_3}{x_1}\gamma_\nu\gamma_\mu\)\Gamma\]\delta(y-x_3)
-2g_{\mu\nu}\Tr\[\overline{\Gamma}^+_k\Gamma\]\delta(y+x_1)
\Big)+\mathcal{O}(a_s^2),
\\
&&S_{\mu\nu,\alpha\beta}^{\Gamma}(x_{1,2,3},y)=
\\\nn && \qquad\qquad
-\frac{a_s(\mu)}{4}\(\theta(x_1,x_3)-\theta(-x_1,-x_3)\)\delta(y+x_2)
g_{\mu\alpha}\Tr\[\(\frac{x_1}{x_2}\gamma_\nu \gamma_\beta-\frac{x_3}{x_2}\gamma_\beta \gamma_\nu\)\gamma^-\Gamma\]+\mathcal{O}(a_s^2).
\end{eqnarray}
Here, the function $\widehat{\Phi}_{\mu,\bullet}$ is regular at $b\to 0$, and represents the effects of pure twist-three dynamics. 

The computation for anti-quark distribution is identical up to a few signs. Here, for completeness, we present the final results. The small-$b$ expansion for anti-quark distributions for the quark channel are
\begin{eqnarray}
\label{matching:12-anti}
\overline{\Phi}^{[\Gamma]}_{\mu,12}(x_{1,2,3},b)\Big|_{\text{quark}}&=&-a_s(\mu)\frac{C_F}{2}\frac{b^\nu}{b^2}\(\theta(x_1,x_2)-\theta(-x_1,-x_2)\)
\\\nn &&
\times
\sum_k
\Tr\[\Gamma\(\gamma_\nu\gamma_\mu-\frac{x_1}{x_3}\gamma_\mu\gamma_\nu\)\overline{\Gamma}^+_k\]
\overline{\phi}^{[\Gamma^+_k]}(x_3,\mu)+\mathcal{O}\(b^0,\frac{a_s^2}{b}\),
\\\label{matching:21-anti}
\overline{\Phi}^{[\Gamma]}_{\mu,21}(x_{1,2,3},b)\Big|_{\text{quark}}&=&-a_s(\mu)\frac{C_F}{2}\frac{b^\nu}{b^2}\(\theta(x_2,x_3)-\theta(-x_2,-x_3)\)
\\\nn &&
\times
\sum_k
\Tr\[\overline{\Gamma}^+_k\(\gamma_\mu\gamma_\nu-\frac{x_3}{x_1}\gamma_\nu\gamma_\mu\)\Gamma\]
\overline{\phi}^{[\Gamma^+_k]}(-x_1,\mu)+\mathcal{O}\(b^0,\frac{a_s^2}{b}\).
\end{eqnarray}
In these equations, the subtraction procedure, eq.~(\ref{def:subtraction}), is already applied. The gluon channel reads
\begin{eqnarray}\nn
\overline{\Phi}^{[\Gamma]}_{\mu,12}(x_{1,2,3},b)\Big|_{\text{gluon}}&=&
\overline{\Phi}^{[\Gamma]}_{\mu,21}(x_{1,2,3},b)\Big|_{\text{gluon}}
\\\label{matching:12-anti-gluon}
&=&
-\frac{a_s(\mu)}{4}\frac{b_\nu}{b^2}\(\theta(x_1,x_3)-\theta(-x_1,-x_3)\)
\\\nn &&
\times
\Tr\[\(\frac{x_3}{x_2}\gamma^\nu\gamma^\alpha -\frac{x_1}{x_2}\gamma^\alpha\gamma^\nu\)\gamma^-\Gamma\]
\phi_{\mu\alpha}(-x_2,\mu)+\mathcal{O}\(b^0,\frac{a_s^2}{b}\).
\end{eqnarray}
Combining these formulas together and presenting collinear distributions via the TMD ones we obtain
\begin{eqnarray}
\overline{\mathbf{\Phi}}_{\mu,12}^{[\Gamma]}(x_{1,2,3},b)&=&
\widehat{\overline{\mathbf{\Phi}}}_{\mu,12}^{[\Gamma]}(x_{1,2,3},b)
\\\nn&&
+\frac{b^\nu}{b^2}\int dy\(-\sum_k  S_{\mu\nu}^{\dagger\Gamma,\overline{\Gamma}^+_k}(x_{1,2,3},y) \overline{\Phi}_{11}^{[\Gamma_k^+]}(y,b)
+S_{\mu\nu,\alpha\beta}^{\Gamma}(x_{3,2,1},y) \Phi_{11}^{\alpha\beta}(y,b)\),
\\
\overline{\mathbf{\Phi}}_{\mu,21}^{[\Gamma]}(x_{1,2,3},b)&=&
\widehat{\overline{\mathbf{\Phi}}}_{\mu,21}^{[\Gamma]}(x_{1,2,3},b)
\\\nn&&
+\frac{b^\nu}{b^2}\int dy\(\sum_k  S_{\mu\nu}^{\Gamma,\overline{\Gamma}^+_k}(-x_{3,2,1},y) \Phi_{11}^{[\Gamma_k^+]}(y,b)+S_{\mu\nu,\alpha\beta}^{\Gamma}(x_{3,2,1},y) \Phi_{11}^{\alpha\beta}(y,b)\).
\end{eqnarray}

It should be taken into account that only TMD distributions that posses a matching to the collinear twist two contribute on the right-hand side of L-TMD approximation. For example, the Boer-Mulders function should not be included, because it matches to the twist-three collinear operator \cite{Scimemi:2018mmi}, and thus could not appear in eq.~(\ref{matching:12-sing}-\ref{matching:21-sing}). 

The  L-TMD approximation can be systematically improved with higher perturbative orders. The only missing ingredient in the results  shown in this work is the NLO coefficient function for the singular part. It is expected that at NLO, the L-TMD approximation is more involved due to the mixing of flavors within OPE. 

\subsection{L-TMD approximation for TMD distributions}
\label{sec:L-TMD-LA}

The cross-section at NLP involves TMD distributions $\mathbf f_\bullet^\perp$, $\mathbf g_\bullet^\perp$, and $\mathbf h_\bullet$.  To obtain the L-TMD approximation for these function we should compute the traces presented in eq.~(\ref{def:S1}) and compare the parameterizations. The functions $\mathbf h_{\oplus}$ and $\mathbf h_{\ominus}$ do not have any singularity at small-$b$, and thus are zero in the L-TMD approximation. For the functions $\mathbf f_{\ominus}^\perp$ and $\mathbf g_{\oplus}^\perp$ we obtain\footnote{In total there are 8 twist-three distributions which have singular starting at LO, and another 8 which have singular matching starting at NLO. The list of these distributions is given in appendix C.3 of ref.~\cite{Rodini:2022wki}}
\begin{eqnarray}
\mathbf f_{\ominus}^\perp\Big|_{\text{L-TMD}}&=&
\frac{1}{b^2 M^2}\Big[2a_sC_F \(\theta(x_2,x_3)-\theta(-x_2,-x_3)\)\(\frac{x_1-x_3}{x_1}f_1(-x_1,b)-2f_1(x_3,b)\)
\\\nn && +\frac{a_s}{2}\frac{x_1-x_3}{x_2}\(\theta(x_1,x_3)-\theta(-x_1,-x_3)\)f_g(-x_2,b)+\mathcal{O}(a_s^2)\Big],
\\
\mathbf g_{\oplus}^\perp\Big|_{\text{L-TMD}}&=&\frac{1}{b^2 M^2}\Big[2a_sC_F\(\theta(x_2,x_3)-\theta(-x_2,-x_3)\)\frac{x_2}{x_1}f_1(-x_1,b)
\\\nn && +\frac{a_s}{2}\(\theta(x_1,x_3)-\theta(-x_1,-x_3)\)f_g(-x_2,b)+\mathcal{O}(a_s^2)\Big].
\end{eqnarray}
The L-TMD approximation for anti-quark distributions is
\begin{eqnarray}
\overline{\mathbf f}_{\ominus}^\perp\Big|_{\text{L-TMD}}&=&
\frac{-1}{b^2 M^2}\Big[2a_sC_F \(\theta(x_2,x_3)-\theta(-x_2,-x_3)\)\(\frac{x_1-x_3}{x_1}\overline{f}_1(-x_1,b)-2\overline{f}_1(x_3,b)\)
\\\nn && +\frac{a_s}{2}\frac{x_1-x_3}{x_2}\(\theta(x_1,x_3)-\theta(-x_1,-x_3)\)f_g(-x_2,b)+\mathcal{O}(a_s^2)\Big],
\\
\overline{\mathbf g}_{\oplus}^\perp\Big|_{\text{L-TMD}}&=&\frac{1}{b^2 M^2}\Big[2a_sC_F \(\theta(x_2,x_3)-\theta(-x_2,-x_3)\)
\frac{x_2}{x_1}\overline{f}_1(-x_1,b)
\\\nn && +\frac{a_s}{2}\(\theta(x_1,x_3)-\theta(-x_1,-x_3)\)f_g(-x_2,b)+\mathcal{O}(a_s^2)\Big].
\end{eqnarray}
The distributions $\mathbf f_{\oplus}^\perp$ and $\mathbf g_{\ominus}^\perp$ do not have a singular matching at LO, but it appears at NLO order (two-loops). Note, that for the DY cross-section in eq.~(\ref{W_gNLP}) the variable $x_3>0$. Consequently, the arguments of twist-two TMD distributions are always positive.

The combinations $\mathbf{f}_2^\perp$ that appear in the cross-section at NLP eq.~(\ref{combinations}) in L-TMD approximation take the form
\begin{eqnarray}\label{L-TMD:f2}
\mathbf{f}_2^\perp\Big|_{\text{L-TMD}}&=&
\frac{1}{b^2 M^2}\Big[
4a_sC_F \(\theta(x_2,x_3)-\theta(-x_2,-x_3)\)\(f_1(-x_1,b)-f_1(x_3,b)\)
\\\nn &&
+a_s\frac{x_1}{x_2}\(\theta(x_1,x_3)-\theta(-x_1,-x_3)\)f_g(-x_2,b)+\mathcal{O}(a_s^2)\Big],
\\\label{L-TMD:f2-bar}
\overline{\mathbf{f}}_2^\perp\Big|_{\text{L-TMD}}&=&
\frac{-1}{b^2 M^2}\Big[
4a_sC_F \(\theta(x_2,x_3)-\theta(-x_2,-x_3)\)\(\overline{f}_1(-x_1,b)-\overline{f}_1(x_3,b)\)
\\\nn &&
+
a_s\frac{x_1}{x_2}\(\theta(x_1,x_3)-\theta(-x_1,-x_3)\)f_g(-x_2,b)+\mathcal{O}(a_s^2)\Big],
\end{eqnarray}
The twist-three function $\mathbf{g}_2^\perp$ is suppressed by an extra power of $a_s$
\begin{eqnarray}
\mathbf{g}_2^\perp\Big|_{\text{L-TMD}}=\frac{1}{b^2M^2}\times\mathcal{O}(a_s^2).
\end{eqnarray}

\subsection{Angular distributions in L-TMD approximation}
\label{sec:L-TMD-AS}

Finally, we can report the genuine part of the angular coefficients in the L-TMD approximation. Since only the distribution $\mathbf{f}_2^\perp$ is non-zero at LO,  most part of eq.~(\ref{gNLP:1M}-\ref{gNLP:7M}) is vanishing. Furthermore, the LO convolutions with $\mathbb{C}_I$ is also vanishing, because $\mathbb{C}_I\sim \delta(u_2)$, while the quark part of $\mathbf{f}_2^\perp$ is vanishing at this point, and the gluon part is non-zero only for $u_2<-x<0$. Consequently, only the  convolutions of type $\mathcal{J}_R$ with distributions $\mathbf{f}_2^\perp$ produce non-zero contribution. We find
\begin{eqnarray}\label{L-TMD:1}
\Sigma_1^{\text{L-TMD}}&=&\frac{8\pi \alpha_{\text{em}}^2}{3N_cs Q^2}\sum_{f,G,G'}Q^4\Delta_G^*\Delta_{G'}
\frac{-2|\vec q_T|}{Q}
z_{+\ell}^{GG'}z_{+f}^{GG'}\mathcal{J}^{\text{L-TMD}}_+
\\
\Sigma_3^{\text{L-TMD}}&=&\frac{8\pi \alpha_{\text{em}}^2}{3N_cs Q^2}\sum_{f,G,G'}Q^4\Delta_G^*\Delta_{G'}
\frac{-4|\vec q_T|}{\sqrt{\tau^2}}z_{-\ell}^{GG'}z_{-f}^{GG'}\mathcal{J}^{\text{L-TMD}}_-
,
\\\label{L-TMD:6}
\Sigma_6^{\text{L-TMD}}&=&\mathcal{O}(a_s^2),
\\\label{L-TMD:7}
\Sigma_7^{\text{L-TMD}}&=&\mathcal{O}(a_s^2).
\end{eqnarray}
The convolutions $\mathcal{J}_\pm^{\text{L-TMD}}$ are defined as 
\begin{eqnarray}\label{L-TMD-convolution-PM}
\mathcal{J}^{\text{L-TMD}}_\pm&=&\frac{1}{|\vec q_T|}\int_0^\infty \frac{db}{2\pi} J_1(b|\vec q_T|)\(\frac{\tau^2}{\zeta_\mu}\)^{-2\mathcal{D}(b,\mu)}
\Big[
\\\nn &&
s_{qq}\otimes f_{1}(x_1,b)\overline{f}_{1}(x_2,b)
\pm s_{qq}\otimes \overline{f}_{1}(x_1,b)f_{1}(x_2,b)
\\\nn &&
-f_{1}(x_1,b)s_{qq}\otimes \overline{f}_{1}(x_2,b)
\mp \overline{f}_{1}(x_1,b)s_{qq}\otimes f_{1}(x_2,b)
\\\nn &&
+s_{qg}\otimes f_{g}(x_1,b)\(f_{1}(x_2,b)\pm \overline{f}_{1}(x_2,b)\)
\\\nn &&
-\(f_{1}(x_1,b)\pm\overline{f}_{1}(x_1,b)\)s_{qg}\otimes f_{g}(x_2,b)
\Big],
\end{eqnarray}
where $\otimes$ indicates the ordinary Mellin convolution
\begin{eqnarray}
s\otimes f(x)=\int_x^1 \frac{dy}{y}s(y)f\(\frac{x}{y}\).
\end{eqnarray}
The coefficient functions $s_{qq}$ and $s_{qg}$ at LO are
\begin{eqnarray}
s_{qq}(x)&=&\frac{4a_sC_F}{(1-x)_+}+\mathcal{O}(a_s^2),\qquad
s_{qg}(x)=a_s(1-x)+\mathcal{O}(a_s^2).
\end{eqnarray}
We emphasize that the Collins-Soper kernel in eq.~(\ref{L-TMD-convolution-PM}) is for the quark flavor, despite one of TMD distributions is the gluon TMD distribution.

In eq.~(\ref{L-TMD:1}) and (\ref{L-TMD:7}) we intentionally extracted the factor $|\vec q_T|$, because it properly represents the behavior of this power correction. The convolutions $\mathcal{J}_\pm^{\text{L-TMD}}$ are dimensionless, finite at $q_T\to0$ and have a typical LP behavior,  despite of the common factor $|\vec q_T|^{-1}$. This can be seen using a well-known identity for Bessel functions
\begin{eqnarray}
\int_0^\infty \frac{db}{2\pi |\vec q_T|} J_1(b|\vec q_T|) F(b)=
\frac{1}{2}\int_0^\infty \frac{b\,db}{2\pi} (J_0(b|\vec q_T|)+J_2(b|\vec q_T|)) F(b),
\end{eqnarray}
where $F(b)$ is a decreasing scalar function. The right-hand-side of this relation is finite at $q_T\to0$, and is structurally similar to the LP convolutions eq.~(\ref{def:Jn-S}). Therefore, we explicitly confirm our original expectation that terms extracted by the L-TMD approximation are responsible for the generation of $q_T/Q$ correction.

It is instructive to compare eq.~(\ref{L-TMD:1}-\ref{L-TMD:7}) with the existing collinear computations of ref.~\cite{Mirkes:1994eb, Boer:2006eq, Lyubovitskij:2024civ, Lyubovitskij:2024jlb}. The comparison cannot be done one-to-one, because both factorizations are frame-dependent and describe different kinematical limits, and therefore cannot be matched. Moreover, some parts are hidden in the KPC. Nonetheless, we observe the same combinations of flavors and integral convolutions as in eq.~(\ref{L-TMD-convolution-PM}), which provide an extra check of algebra. The distributions $\Sigma_{6,7}$ are also predicted by collinear factorization as $\mathcal{O}(a_s^2)$ \cite{Lyubovitskij:2024civ}, and also obey the same composition of distributions that follows from L-TMD approximation.

\begin{figure}[t]
\centering
\includegraphics[width=0.99\textwidth]{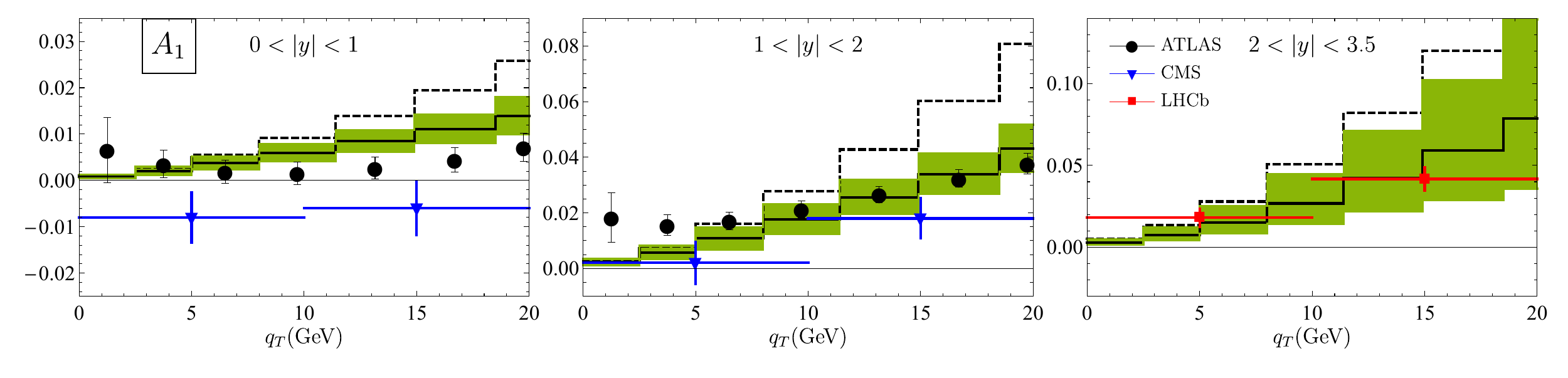}
\caption{\label{fig:A1} Comparison of $A_1$ measurement at $Q\in[80,100]$GeV by ATLAS \cite{ATLAS:2016rnf}, CMS \cite{CMS:2015cyj} and LHCb \cite{LHCb:2022tbc} during the $\sqrt{s}=8$TeV run to the theoretical prediction. The dashed line shows the prediction of KPC term computed in ref.~\cite{Piloneta:2024aac} (shown without uncertainties). The solid line demonstrates the sum of KPC and L-TMD part. The computation is done using ART23 \cite{Moos:2023yfa} TMDPDFs and their uncertainty (green band).}
\end{figure}

The comparison of the theoretical prediction with the data by LHC \cite{ATLAS:2016rnf, CMS:2015cyj, LHCb:2022tbc} is shown in fig.~\ref{fig:A1}. The theoretical prediction includes the KPC term and the $q_T/Q$ term. The unpolarized TMDPDF are taken from ART23 extraction \cite{Moos:2023yfa}. The non-perturbative part for the optimal gluon TMDPDF, which is not known, is taken as $\exp(-\vec b^2/2)$ and we find little sensitivity to this function. We also show the resummed-KPC term  determined in ref.~\cite{Piloneta:2024aac} by a dashed line. One can clearly see that the KPC grows faster than the data. The correction provided by $q_T/Q$ term has negative sign and compensates this growth. Together these terms provide a very good agreement with the data.

\begin{figure}[t]
\centering
\includegraphics[width=0.99\textwidth]{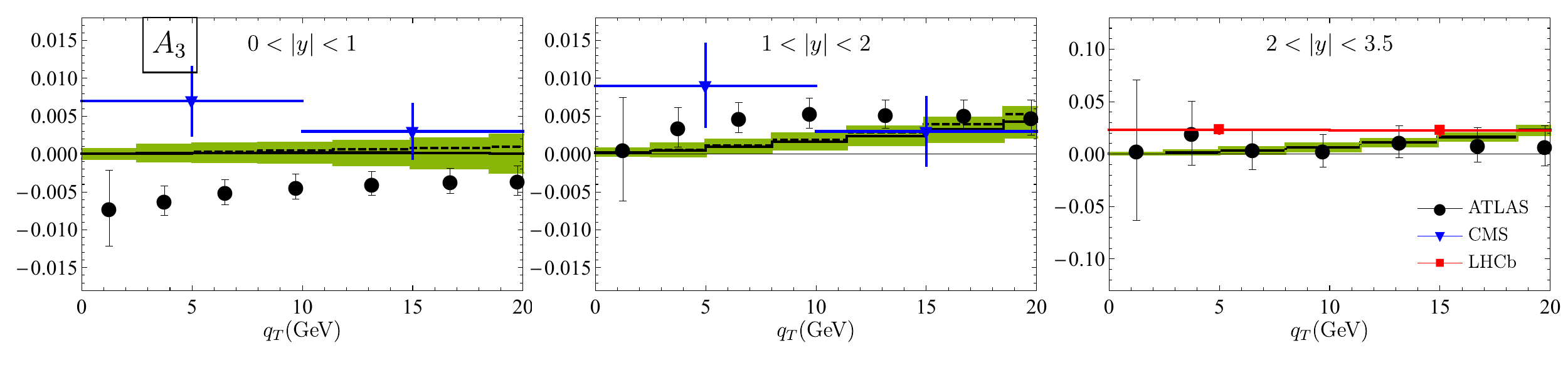}
\caption{\label{fig:A3} Comparison of $A_3$ measurement at $Q\in[80,100]$GeV by ATLAS \cite{ATLAS:2016rnf}, CMS \cite{CMS:2015cyj} and LHCb \cite{LHCb:2022tbc} during the $\sqrt{s}=8$TeV run to the theoretical prediction. The dashed line shows the prediction of KPC term computed in ref.~\cite{Piloneta:2024aac} (shown without uncertainties). The solid line demonstrates the sum of KPC and L-TMD part. The computation is done using ART23 \cite{Moos:2023yfa} TMDPDFs and their uncertainty (green band). }
\end{figure}

The comparison with experimental measurements of $A_3$ is shown in fig.~\ref{fig:A3}. This distribution is very small, mainly due to the $z_{-\ell}$ constant which is $z_{-\ell}^{ZZ}\approx -0.05$. Alike $A_1$, the sizes of KPC and $q_T/Q$ corrections are similar at $q_T\sim 10$-15GeV. The theoretical prediction is in agreement with the data, although the later is a bit controversial and uncertain.

\section{Conclusion}

In this work, we suggest an approach for the explicit determination of $q_T/Q$-corrections within TMD factorization. It consists in isolating singular terms and approximating them in terms of LP TMD distributions. This approximation, which we call the leading-TMD (L-TMD) approximation, preserves the original evolution properties of TMD distributions and thus maintains the consistency of the TMD factorization theorem. The drawback is the redundant freedom in the definition of higher-twist distributions, which becomes dependent on the explicit form of the L-TMD approximation. Nonetheless, this ambiguity is relevant only at low-$Q$ and is therefore marginal compared to the significant improvement in the predictive power of the TMD factorization approach.

As a study case, we consider the L-TMD approximation for the angular distributions of the Drell-Yan lepton pair at NLP. Among eight distributions, the linear $q_T/Q$ correction appears only in the angular distributions $A_{1,3,6,7}$. The L-TMD approximation describes all of them, but for $A_6$ and $A_7$, it requires an NLO perturbative computation. For $A_{1}$ and $A_3$ we find a very good agreement between experimental measurements and the theoretical predictions computed here. Let us also mention that, in this work, for the first time, we present the DY angular distributions at NLP with a clear separation of different twist contributions and including NLO coefficient functions. 

The approximation that we suggest opens a new avenue for studies of TMD factorization framework. The L-TMD approximation can be formulated for the TMD distributions of higher-TMD-twist too. Then, one can consider higher-power terms of  the factorization and extract their $q_T/Q$ part. Ultimately, one may think about an all-power resummation with the L-TMD approximation, which can lead to the description of the whole-range $q_T$-spectrum within the TMD approach.

\acknowledgments

A.V. is funded by the \textit{Atracci\'on de Talento Investigador} program of the Comunidad de Madrid (Spain) No. 2020-T1/TIC-20204.  The project is supported by grants ``Europa Excelencia'' No. EUR2023-143460 funded by MCIN/AEI/10.13039/501100011033/ by the Spanish Ministerio de Ciencias y Innovaci\'on and No. PID2022-136510NB-C31 funded by MCIN/AEI/10.13039/501100011033 by the Spanish Ministerio de Ciencias y Innovaci\'on. This project is also supported by the European Union Horizon research Marie Skłodowska-Curie Actions – Staff Exchanges, HORIZON-MSCA-2023-SE-01-101182937-HeI, DOI: 10.3030/101182937.

\appendix

\section{Lepton tensor decomposition}
\label{app:lepton-tensor}

The angular sub-structure of the cross-section is entirely encoded in the lepton tensor. For that reason, it is most convenient to decompose the lepton tensor priory to the convolution with the hadron tensor in the system of initial hadrons. The corresponding decomposition has been derived in ref.~\cite{Piloneta:2024aac}. For completeness we summarize it here.

The Born approximation for the unpolarized lepton tensor (induced by the vector bosons $G$ and $G'$) reads
\begin{eqnarray}\label{L_0}
L^{\mu\nu}_{GG'}&=&4\[z_{+\ell}^{GG'}(l^\mu l'^\nu+l'^\mu l^\nu-g^{\mu\nu}(ll'))
-iz_{-\ell}^{GG'} \epsilon^{\mu\nu\alpha\beta}l_\alpha l_\beta' \].
\end{eqnarray}
Here, $l$ and $l'$ are the momentum of the negatively and positively charged leptons, respectively. The factors $z^{GG'}_{\pm\ell}$ denote the subsequent electro-weak coupling combinations
\begin{eqnarray}
z^{GG'}_{+\ell}&=&2(g^R_Gg^R_{G'}+g^L_Gg^L_{G'})=\frac{v^Gv^{G'}+a^{G}a^{G'}}{4s_W^2c_W^2},
\\
z^{GG'}_{-\ell}&=&2(g^R_Gg^R_{G'}-g^L_Gg^L_{G'})=-\frac{v^Ga^{G'}+a^{G}v^{G'}}{4s_W^2c_W^2}, 
\end{eqnarray}
where the subscript $\ell$ is used to indicate that these coupling constants are taken with lepton quantum numbers.

The decomposition of this tensor with respect to azimutal angles of leptons in the Collins-Soper frame \cite{Collins:1977iv} is
\begin{eqnarray}
L^{\mu\nu}_{GG'}&=&(-Q^2)\(
z^{GG'}_{+\ell}\sum_{n=U,0,1,2,5,6}S_n(\theta,\phi)\mathfrak{L}_n^{\mu\nu}
+
z^{GG'}_{-\ell}\sum_{n=3,4,7}S_n(\theta,\phi)\mathfrak{L}_n^{\mu\nu}\),
\end{eqnarray}
where the $S_n(\theta,\phi)$ variables constitute a set of independent angular structures and $\mathfrak{L}_n^{\mu\nu}$ are known tensors. These angular structures read
\begin{align}\nn
&S_{U}(\theta,\phi)=1+\cos^2\theta,
&&S_0(\theta,\phi)=\frac{1-3\cos^2\theta}{2},
&&S_1(\theta,\phi)=\sin2\theta\cos\phi,
\\\label{S-def}
&S_2(\theta,\phi)=\frac{1}{2}\sin^2\theta\cos2\phi
&&S_3(\theta,\phi)=\sin\theta \cos\phi,
&&S_4(\theta,\phi)=\cos\theta,
\\\nn
&S_5(\theta,\phi)=\sin^2\theta\sin2\phi
&&S_6(\theta,\phi)=\sin2\theta \sin\phi,
&&S_7(\theta,\phi)=\sin\theta\sin\phi.
\end{align}
Upon the integration over the full solid angle, all structures, except $S_{U}$, vanish
\begin{eqnarray}
\int d\Omega\, S_n(\theta,\phi)=
\left\{\begin{array}{cl}
\Ds\frac{16\pi}{3},& n=U,
\\
0,&\text{otherwise}.
\end{array}\right.
\end{eqnarray}
The $\mathfrak{L}_n^{\mu\nu}$ tensors are
\begin{eqnarray}\label{L:U}
\mathfrak{L}^{\mu\nu}_{U}&=&g^{\mu\nu}-\frac{q^\mu q^\nu}{Q^2},
\\\label{L:0}
\mathfrak{L}^{\mu\nu}_{0}&=&-q_+q_-\(\frac{n^\mu}{q_+}-\frac{\bar n^\mu}{q_-}\)\(\frac{n^\nu}{q_+}-\frac{\bar n^\nu}{q_-}\),
\\\label{L:1}
\mathfrak{L}^{\mu\nu}_{1}&=&q_+q_-\frac{Q}{|\vec q_T|}
\[
\(\frac{n^\mu}{q_+}-\frac{q^\mu}{Q^2}\)\(\frac{n^\nu}{q_+}-\frac{q^\nu}{Q^2}\)
-
\(\frac{\bar n^\mu}{q_-}-\frac{q^\mu}{Q^2}\)\(\frac{\bar n^\nu}{q_-}-\frac{q^\nu}{Q^2}\)\]
,
\\\label{L:2}
\mathfrak{L}^{\mu\nu}_{2}&=&2\(g^{\mu\nu}-\frac{q^\mu q^\nu}{Q^2}\)+\(2\frac{Q^2}{\vec q_T^2}-1\)\mathfrak{L}^{\mu\nu}_{0}
\\\nn &&+
4\frac{Q^2}{\vec q_T^2}q_+q_-\[
\(\frac{n^\mu}{q_+}-\frac{q^\mu}{Q^2}\)\(\frac{n^\nu}{q_+}-\frac{q^\nu}{Q^2}\)
+
\(\frac{\bar n^\mu}{q_-}-\frac{q^\mu}{Q^2}\)\(\frac{\bar n^\nu}{q_-}-\frac{q^\nu}{Q^2}\)\]
,
\\\label{L:3}
\mathfrak{L}^{\mu\nu}_{3}&=&i\frac{\tau}{|\vec q_T|}\(\frac{\epsilon^{\mu\nu \alpha \beta}q_\alpha n_\beta}{q_+}+\frac{\epsilon^{\mu\nu \alpha \beta}q_\alpha \bar n_\beta}{q_-}\)
\\\label{L:4}
\mathfrak{L}^{\mu\nu}_{4}&=&i\frac{\tau}{Q}\(\frac{\epsilon^{\mu\nu \alpha \beta}q_\alpha n_\beta}{q_+}-\frac{\epsilon^{\mu\nu \alpha \beta}q_\alpha \bar n_\beta}{q_-}\)
,
\\\label{L:5}
\mathfrak{L}^{\mu\nu}_{5}&=&\frac{-\tau Q}{2\vec q^2_T}\[
\(2\frac{q^\mu}{Q^2}-\frac{n^\mu}{q_+}-\frac{\bar n^\mu}{q_-}\)\tilde q^\nu
+
\tilde q^\mu\(2\frac{q^\nu}{Q^2}-\frac{n^\nu}{q_+}-\frac{\bar n^\nu}{q_-}\)\],
\\\label{L:6}
\mathfrak{L}^{\mu\nu}_{6}&=&\frac{\tau}{2|\vec q_T|}\[
\(\frac{n^\mu}{q_+}-\frac{\bar n^\mu}{q_-}\)\tilde q^\nu
+
\tilde q^\mu\(\frac{n^\nu}{q_+}-\frac{\bar n^\nu}{q_-}\)\],
\\\label{L:7}
\mathfrak{L}^{\mu\nu}_{7}&=&
-i \frac{Q\tau^2}{|\vec q_T|}
\[
\(\frac{n^\mu}{q_+}-\frac{q^\mu}{q^2}\)\(\frac{\bar n^\nu}{q_-}-\frac{q^\nu}{q^2}\)
-
\(\frac{\bar n^\mu}{q_-}-\frac{q^\mu}{q^2}\)\(\frac{n^\nu}{q_+}-\frac{q^\nu}{q^2}\)
\],
\end{eqnarray}
where the vectors $\bar n$ and $n$ are defined by hadron's momenta $p_1$ and $p_2$, correspondingly, and 
\begin{eqnarray}
\tilde q_T^\mu =\epsilon^{\mu\nu \rho \lambda}q_\nu \bar n_\rho n_\lambda.
\end{eqnarray}
The tensors $\mathfrak{L}$ are dimensionless and transverse to $q^\mu$, i.e., $q_\mu \mathfrak{L}^{\mu\nu}_n=q_\nu \mathfrak{L}^{\mu\nu}_n=0$.

\section{Coupling constants and propagators for neutral bosons}
\label{app:couplings}

The computation of Z/$\gamma$-boson production involves the current
\begin{eqnarray}
\omega_{G}^\mu =g_R^{G}\gamma^\mu(1+\gamma^5)+g_L^{G}\gamma^\mu(1-\gamma^5)=\frac{1}{2s_Wc_W}\gamma^\mu (v_G-a_G\gamma^5),
\end{eqnarray}
where $g_R^G$, $g_L^G$, $v_G$, and $a_G$ represent the right, left, vector and axial electroweak coupling constants, respectively, and the label $G$ denotes the type of gauge boson. The left/right and vector/axial couplings relate to each other as
\begin{eqnarray}
v_f=2s_Wc_W(g_{R}^{G}+g_{L}^{G}),\qquad
a_f=2s_Wc_W(g_{L}^{G}-g_{R}^{G}),
\end{eqnarray}
where $s_W$ and $c_W$ are sine and cosine of Weinberg angle. The couplings $g_R^G$ and $g_L^G$, particularized for neutral electroweak bosons , are
\begin{eqnarray}\label{explicit_g}
g_R^Z=\frac{-e_fs_W^2}{2s_Wc_W},\qquad g_L^Z=\frac{T_3-e_fs_W^2}{2s_Wc_W},\qquad g_\gamma^R=g_\gamma^L=\frac{e_f}{2},
\end{eqnarray}
where $e_f$ is the electric charge of a fermion $f$ (in units of $e$), $T_3$ is the third projection of its weak isospin.

Describing the Drell-Yan reaction the couplings appear in the following combinations
\begin{eqnarray}\label{def:zP}
z_{+}^{GG'}&=&2(g_{R}^{G}g_{R}^{G'}+g_{L}^{G}g_{L}^{G'})=\frac{v^G v^{G'}+a^G a^{G'}}{4 s_W^2c_W^2},
\\ \label{def:rP}
r_{+}^{GG'}&=&2(g_{R}^{G}g_{L}^{G'}+g_{L}^{G}g_{R}^{G'})=\frac{v^G v^{G'}-a^G a^{G'}}{4 s_W^2c_W^2},
\\ \label{def:zM}
z_{-}^{GG'}&=&2(g_{R}^{G}g_{R}^{G'}-g_{L}^{G}g_{L}^{G'})=-\frac{v^G a^{G'}+a^G v^{G'}}{4 s_W^2c_W^2},
\\ \label{def:rM}
r_{-}^{GG'}&=&2(g_{R}^{G}g_{L}^{G'}-g_{L}^{G}g_{R}^{G'})=\frac{v^G a^{G'}-a^G v^{G'}}{4 s_W^2c_W^2}.
\end{eqnarray}
All possible coupling constants combinations for the neutral bosons we are interested in can be obtained by substituting eq.~(\ref{explicit_g}) into eq.~(\ref{def:zP}-\ref{def:rM}) 
\begin{eqnarray}
&&z_{+f}^{\gamma\gamma}=r_{+f}^{\gamma\gamma}=e_f^2,
\\\nn
&&z_{-f}^{\gamma\gamma}=r_{-f}^{\gamma\gamma}=0,
\\
&&z_{+f}^{\gamma Z}=z_{+f}^{Z\gamma}=r_{+f}^{\gamma Z}=r_{+f}^{Z\gamma}=\frac{e_f(T_3-2e_fs_W^2)}{2s_Wc_W} = \frac{|e_f|-4e_f^2s_W^2}{4s_wc_W},
\\
&&z_{+f}^{ZZ}=\frac{T_3^2-2e_fT_3s_W^2+2e_f^2s_W^4}{2s^2_Wc^2_W}=\frac{(1-2|e_f|s_W^2)^2+4e_f^2s_W^4}{8s_W^2c_W^2},
\\
&&r_{+f}^{ZZ}=\frac{e_f(e_fs_W^2-T_3)}{c_W^2}=\frac{2e^2_fs^2_W-|e_f|}{2c_W^2},
\\
&&z_{-f}^{\gamma Z}=z_{-f}^{Z\gamma}=-r_{-f}^{\gamma Z}=r_{-f}^{Z\gamma}=-\frac{T_3e_f}{2s_Wc_W}=-\frac{|e_f|}{4s_Wc_W},
\\
&&z_{-f}^{ZZ}=\frac{T_3(2e_fs_W^2-T_3)}{2s_W^2c_W^2}=\frac{4|e_f|s_W^2-1}{8s_W^2c_W^2},
\\
&& r_{-f}^{ZZ}=0.
\end{eqnarray}
Note, that $f$ denotes the flavor of the quark fields. For the case of lepton, values of $e_f$ and $T_3$ must be replaced by corresponding values for the lepton ($-1$ and $-1/2$ correspondingly).

We also list the sums of the gauge-boson propagators, which appear in the cross-section. Using the neutral boson propagator in the form
\begin{eqnarray}
\Delta_G = \frac{\delta_{G\gamma}}{Q^2+i0} + \frac{\delta_{GZ}}{Q^2-M_Z^2 + i\Gamma_ZM_Z},
\end{eqnarray}
we obtain the list of all appearing combinations is
\begin{eqnarray}
&&\sum_{GG'}Q^4z_{+\ell}^{GG'}z_{+f}^{GG'}\Delta^*_G\Delta_{G'}=
\\\nn &&
\qquad
z_{+\ell}^{\gamma\gamma}z_{+f}^{\gamma\gamma}
+
z_{+\ell}^{\gamma Z}z_{+f}^{\gamma Z}\frac{2Q^2(Q^2-M_Z^2)}{(Q^2-M_Z^2)^2+\Gamma_Z^2M_Z^2}
+
z_{+\ell}^{Z Z}z_{+f}^{Z Z}\frac{Q^4}{(Q^2-M_Z^2)^2+\Gamma_Z^2M_Z^2},
\\
&&\sum_{GG'}Q^4z_{+\ell}^{GG'}r_{+f}^{GG'}\Delta^*_G\Delta_{G'}=
\\\nn &&
\qquad
z_{+\ell}^{\gamma\gamma}z_{+f}^{\gamma\gamma}
+
z_{+\ell}^{\gamma Z}z_{+f}^{\gamma Z}\frac{2Q^2(Q^2-M_Z^2)}{(Q^2-M_Z^2)^2+\Gamma_Z^2M_Z^2}
+
z_{+\ell}^{Z Z}r_{+f}^{Z Z}\frac{Q^4}{(Q^2-M_Z^2)^2+\Gamma_Z^2M_Z^2},
\\
&&\sum_{GG'}Q^4z_{-\ell}^{GG'}z_{-f}^{GG'}\Delta^*_G\Delta_{G'}=
\\\nn &&\qquad
z_{-\ell}^{\gamma Z}z_{-f}^{\gamma Z}\frac{2Q^2(Q^2-M_Z^2)}{(Q^2-M_Z^2)^2+\Gamma_Z^2M_Z^2}
+
z_{-\ell}^{Z Z}z_{-f}^{Z Z}\frac{Q^4}{(Q^2-M_Z^2)^2+\Gamma_Z^2M_Z^2},
\\\label{app:interf-term}
&&\sum_{GG'}Q^4 iz_{+\ell}^{GG'}r_{-f}^{GG'}\Delta^*_G\Delta_{G'}=
z_{+\ell}^{\gamma Z}r_{-f}^{\gamma Z}\frac{2Q^2 M_Z\Gamma_Z}{(Q^2-M_Z^2)^2+\Gamma_Z^2M_Z^2}.
\end{eqnarray}

Note, the case of $W$-boson can be easily obtained using that $g_R^W=0$, and 
\begin{eqnarray}
g_L^W=\frac{V_{ff'}}{2s_W},\qquad z_+^{WW}=-z_-^{WW}=\frac{|V_{ff'}|^2}{4s_W^2},\qquad r_\pm^{WW}=0.
\end{eqnarray}
Since the couplings are not diagonal in flavors, the TMD distributions should be summed in the correct flavor combinations.

\bibliography{bibFILE}

\end{document}

%% file: Figures/diagrams1.tex
\begin{tikzpicture}

\coordinate (A) at (1, 2);

\draw[thick,postaction={decorate, decoration={
    markings,
    mark=at position 0.6 with {\arrow{stealth}}}}] ($(A)+(0,-2)$)--($(A)+(0,0)$);

\draw[thick,postaction={decorate, decoration={
    markings,
    mark=at position 0.6 with {\arrow{stealth}}}}] ($(A)+(1.5,-1)$)--($(A)+(1.5,-2)$);    
\draw[thick,postaction={decorate, decoration={
    markings,
    mark=at position 0.6 with {\arrow{stealth}}}}] ($(A)+(1.5,-0.5)$)--($(A)+(1.5,-1)$);    
    
\draw[very thick,style={decorate, decoration={snake,segment length=6.2}}] 
($(A)+(0.5,0)$) to[out=-90,in=-180] ($(A)+(1.5,-1)$);

\draw[thick,dashed] ($(A)$) -- ($(A)+(2.5,0)$);
\draw[thick,dashed] ($(A)+(1.5,-0.5)$) -- ($(A)+(3.5,-.5)$);

\draw[black,fill=black] ($(A)+(0.5,0)$) circle (2pt);
\node[above] at ($(A)+(0.5,0)$) {$F_{\mu+}$};

\node[below] at ($(A)+(0,-2)$) {$q$};
\node[below] at ($(A)+(1.5,-2)$) {$\bar q$};


\coordinate (A) at (5.5, 2);

\draw[thick,postaction={decorate, decoration={
    markings,
    mark=at position 0.6 with {\arrow{stealth}}}}] ($(A)+(0,-2)$)--($(A)+(0,-1)$);

\draw[thick,postaction={decorate, decoration={
    markings,
    mark=at position 0.6 with {\arrow{stealth}}}}] ($(A)+(0,-1)$)--($(A)+(0,0)$);
\draw[thick,postaction={decorate, decoration={
    markings,
    mark=at position 0.8 with {\arrow{stealth}}}}] ($(A)+(1.5,-0.5)$)--($(A)+(1.5,-2)$);    
    
\draw[very thick,style={decorate, decoration={snake,segment length=6.2}}] 
($(A)+(2.5,-0.5)$) to[out=-90,in=0] ($(A)+(0,-1)$);

\draw[thick,dashed] ($(A)$) -- ($(A)+(2.5,0)$);
\draw[thick,dashed] ($(A)+(1.5,-0.5)$) -- ($(A)+(3.5,-.5)$);

\draw[black,fill=black] ($(A)+(2.5,-0.5)$) circle (2pt);
\node[above] at ($(A)+(2.5,-0.5)$) {$F_{\mu+}$};

\node[below] at ($(A)+(0,-2)$) {$q$};
\node[below] at ($(A)+(1.5,-2)$) {$\bar q$};


\coordinate (A) at (10, 2);

\draw[thick,postaction={decorate, decoration={
    markings,
    mark=at position 0.3 with {\arrow{stealth}}}}] ($(A)+(1.5,-1)$)to[out=-180,in=-90]($(A)+(0,0)$);

\draw[very thick,style={decorate, decoration={snake,segment length=6.2}}] ($(A)+(1.5,-1)$)--($(A)+(1.5,-2)$);    
\draw[thick,postaction={decorate, decoration={
    markings,
    mark=at position 0.6 with {\arrow{stealth}}}}] ($(A)+(1.5,-0.5)$)--($(A)+(1.5,-1)$);    
    
\draw[very thick,style={decorate, decoration={snake,segment length=6.2}}] 
($(A)+(0.5,0)$) -- ($(A)+(0.5,-2)$);

\draw[thick,dashed] ($(A)$) -- ($(A)+(2.5,0)$);
\draw[thick,dashed] ($(A)+(1.5,-0.5)$) -- ($(A)+(3.5,-.5)$);

\draw[black,fill=black] ($(A)+(0.5,0)$) circle (2pt);
\node[above] at ($(A)+(0.5,0)$) {$F_{\mu+}$};


\node[below] at ($(A)+(1.5,-2)$) {$A_\nu$};

\coordinate (A) at (14.5, 2);

\draw[very thick,style={decorate, decoration={snake,segment length=6.2}}] ($(A)+(0,-2)$)--($(A)+(0,-1)$);

\draw[thick,postaction={decorate, decoration={
    markings,
    mark=at position 0.6 with {\arrow{stealth}}}}] ($(A)+(0,-1)$)--($(A)+(0,0)$);
\draw[thick,postaction={decorate, decoration={
    markings,
    mark=at position 0.5 with {\arrow{stealth}}}}] ($(A)+(1.5,-0.5)$)to[out=-90,in=0]($(A)+(0,-1)$);    
    
\draw[very thick,style={decorate, decoration={snake,segment length=6.2}}] 
($(A)+(2.5,-0.5)$) --($(A)+(2.5,-2)$);

\draw[thick,dashed] ($(A)$) -- ($(A)+(2.5,0)$);
\draw[thick,dashed] ($(A)+(1.5,-0.5)$) -- ($(A)+(3.5,-.5)$);

\draw[black,fill=black] ($(A)+(2.5,-0.5)$) circle (2pt);
\node[above] at ($(A)+(2.5,-0.5)$) {$F_{\mu+}$};

\node[below] at ($(A)+(0,-2)$) {$A_\nu$};


\end{tikzpicture}

%% file: DY_qTcorrection.bbl
\providecommand{\href}[2]{#2}\begingroup\raggedright\begin{thebibliography}{10}

\bibitem{Collins:2011zzd}
J.~Collins, \emph{{Foundations of perturbative QCD}}, vol.~32, Cambridge
  University Press (11, 2013).

\bibitem{Echevarria:2011epo}
M.G.~Echevarria, A.~Idilbi and I.~Scimemi, \emph{{Factorization Theorem For
  Drell-Yan At Low $q_T$ And Transverse Momentum Distributions
  On-The-Light-Cone}},
  \href{https://doi.org/10.1007/JHEP07(2012)002}{\emph{JHEP} {\bfseries 07}
  (2012) 002} [\href{https://arxiv.org/abs/1111.4996}{{\ttfamily 1111.4996}}].

\bibitem{Becher:2010tm}
T.~Becher and M.~Neubert, \emph{{Drell-Yan Production at Small $q_T$,
  Transverse Parton Distributions and the Collinear Anomaly}},
  \href{https://doi.org/10.1140/epjc/s10052-011-1665-7}{\emph{Eur. Phys. J. C}
  {\bfseries 71} (2011) 1665}
  [\href{https://arxiv.org/abs/1007.4005}{{\ttfamily 1007.4005}}].

\bibitem{Angeles-Martinez:2015sea}
R.~Angeles-Martinez et~al., \emph{{Transverse Momentum Dependent (TMD) parton
  distribution functions: status and prospects}},
  \href{https://doi.org/10.5506/APhysPolB.46.2501}{\emph{Acta Phys. Polon. B}
  {\bfseries 46} (2015) 2501}
  [\href{https://arxiv.org/abs/1507.05267}{{\ttfamily 1507.05267}}].

\bibitem{Boussarie:2023izj}
R.~Boussarie et~al., \emph{{TMD Handbook}},
  \href{https://arxiv.org/abs/2304.03302}{{\ttfamily 2304.03302}}.

\bibitem{Moos:2025sal}
V.~Moos, I.~Scimemi, A.~Vladimirov and P.~Zurita, \emph{{Determination of
  unpolarized TMD distributions from the fit of Drell-Yan and SIDIS data at
  N$^4$LL}},  \href{https://arxiv.org/abs/2503.11201}{{\ttfamily 2503.11201}}.

\bibitem{Bacchetta:2025ara}
{\scshape MAP} collaboration, \emph{{A Neural-Network Extraction of Unpolarised
  Transverse-Momentum-Dependent Distributions}},
  \href{https://arxiv.org/abs/2502.04166}{{\ttfamily 2502.04166}}.

\bibitem{Bacchetta:2024qre}
{\scshape MAP (Multi-dimensional Analyses of Partonic distributions)}
  collaboration, \emph{{Flavor dependence of unpolarized quark transverse
  momentum distributions from a global fit}},
  \href{https://doi.org/10.1007/JHEP08(2024)232}{\emph{JHEP} {\bfseries 08}
  (2024) 232} [\href{https://arxiv.org/abs/2405.13833}{{\ttfamily
  2405.13833}}].

\bibitem{Bacchetta:2024yzl}
A.~Bacchetta, A.~Bongallino, M.~Cerutti, M.~Radici and L.~Rossi,
  \emph{{Exploring the three-dimensional momentum distribution of
  longitudinally polarized quarks in the proton}},
  \href{https://arxiv.org/abs/2409.18078}{{\ttfamily 2409.18078}}.

\bibitem{Yang:2024drd}
K.~Yang, T.~Liu, P.~Sun, Y.~Zhao and B.-Q.~Ma, \emph{{First Extraction of
  Transverse Momentum Dependent Helicity Distributions}},
  \href{https://arxiv.org/abs/2409.08110}{{\ttfamily 2409.08110}}.

\bibitem{Billis:2024dqq}
G.~Billis, J.K.L.~Michel and F.J.~Tackmann, \emph{{Drell-Yan
  transverse-momentum spectra at N$^{3}$LL' and approximate N$^{4}$LL with
  SCETlib}}, \href{https://doi.org/10.1007/JHEP02(2025)170}{\emph{JHEP}
  {\bfseries 02} (2025) 170}
  [\href{https://arxiv.org/abs/2411.16004}{{\ttfamily 2411.16004}}].

\bibitem{Balitsky:2017gis}
I.~Balitsky and A.~Tarasov, \emph{{Power corrections to TMD factorization for
  Z-boson production}},
  \href{https://doi.org/10.1007/JHEP05(2018)150}{\emph{JHEP} {\bfseries 05}
  (2018) 150} [\href{https://arxiv.org/abs/1712.09389}{{\ttfamily
  1712.09389}}].

\bibitem{Balitsky:2020jzt}
I.~Balitsky, \emph{{Gauge-invariant TMD factorization for Drell-Yan hadronic
  tensor at small x}},
  \href{https://doi.org/10.1007/JHEP05(2021)046}{\emph{JHEP} {\bfseries 05}
  (2021) 046} [\href{https://arxiv.org/abs/2012.01588}{{\ttfamily
  2012.01588}}].

\bibitem{Inglis-Whalen:2021bea}
M.~Inglis-Whalen, M.~Luke, J.~Roy and A.~Spourdalakis, \emph{{Factorization of
  power corrections in the Drell-Yan process in EFT}},
  \href{https://doi.org/10.1103/PhysRevD.104.076018}{\emph{Phys. Rev. D}
  {\bfseries 104} (2021) 076018}
  [\href{https://arxiv.org/abs/2105.09277}{{\ttfamily 2105.09277}}].

\bibitem{Balitsky:2021fer}
I.~Balitsky, \emph{{Drell-Yan angular lepton distributions at small x from TMD
  factorization.}}, \href{https://doi.org/10.1007/JHEP09(2021)022}{\emph{JHEP}
  {\bfseries 09} (2021) 022}
  [\href{https://arxiv.org/abs/2105.13391}{{\ttfamily 2105.13391}}].

\bibitem{Vladimirov:2021hdn}
A.~Vladimirov, V.~Moos and I.~Scimemi, \emph{{Transverse momentum dependent
  operator expansion at next-to-leading power}},
  \href{https://doi.org/10.1007/JHEP01(2022)110}{\emph{JHEP} {\bfseries 01}
  (2022) 110} [\href{https://arxiv.org/abs/2109.09771}{{\ttfamily
  2109.09771}}].

\bibitem{Ebert:2021jhy}
M.A.~Ebert, A.~Gao and I.W.~Stewart, \emph{{Factorization for azimuthal
  asymmetries in SIDIS at next-to-leading power}},
  \href{https://doi.org/10.1007/JHEP06(2022)007}{\emph{JHEP} {\bfseries 06}
  (2022) 007} [\href{https://arxiv.org/abs/2112.07680}{{\ttfamily
  2112.07680}}].

\bibitem{Rodini:2022wic}
S.~Rodini and A.~Vladimirov, \emph{{Factorization for quasi-TMD distributions
  of sub-leading power}},
  \href{https://doi.org/10.1007/JHEP09(2023)117}{\emph{JHEP} {\bfseries 09}
  (2023) 117} [\href{https://arxiv.org/abs/2211.04494}{{\ttfamily
  2211.04494}}].

\bibitem{Gamberg:2022lju}
L.~Gamberg, Z.-B.~Kang, D.Y.~Shao, J.~Terry and F.~Zhao,
  \emph{{Transverse-momentum-dependent factorization at next-to-leading
  power}},  \href{https://arxiv.org/abs/2211.13209}{{\ttfamily 2211.13209}}.

\bibitem{Rodini:2023plb}
S.~Rodini and A.~Vladimirov, \emph{{Transverse momentum dependent factorization
  for SIDIS at next-to-leading power}},
  \href{https://doi.org/10.1103/PhysRevD.110.034009}{\emph{Phys. Rev. D}
  {\bfseries 110} (2024) 034009}
  [\href{https://arxiv.org/abs/2306.09495}{{\ttfamily 2306.09495}}].

\bibitem{delCastillo:2023rng}
R.F.~del Castillo, M.~Jaarsma, I.~Scimemi and W.~Waalewijn, \emph{{Transverse
  momentum measurements with jets at next-to-leading power}},
  \href{https://doi.org/10.1007/JHEP02(2024)074}{\emph{JHEP} {\bfseries 02}
  (2024) 074} [\href{https://arxiv.org/abs/2307.13025}{{\ttfamily
  2307.13025}}].

\bibitem{Balitsky:2024ozy}
I.~Balitsky, \emph{{$1/Q^2$ power corrections to TMD factorization for
  Drell-Yan hadronic tensor}},
  \href{https://doi.org/10.1016/j.nuclphysb.2024.116658}{\emph{Nucl. Phys. B}
  {\bfseries 1006} (2024) 116658}
  [\href{https://arxiv.org/abs/2404.15116}{{\ttfamily 2404.15116}}].

\bibitem{Vladimirov:2023aot}
A.~Vladimirov, \emph{{Kinematic power corrections in TMD factorization
  theorem}}, \href{https://doi.org/10.1007/JHEP12(2023)008}{\emph{JHEP}
  {\bfseries 12} (2023) 008}
  [\href{https://arxiv.org/abs/2307.13054}{{\ttfamily 2307.13054}}].

\bibitem{Rodini:2022wki}
S.~Rodini and A.~Vladimirov, \emph{{Definition and evolution of transverse
  momentum dependent distribution of twist-three}},
  \href{https://doi.org/10.1007/JHEP08(2022)031}{\emph{JHEP} {\bfseries 08}
  (2022) 031} [\href{https://arxiv.org/abs/2204.03856}{{\ttfamily
  2204.03856}}].

\bibitem{ATLAS:2016rnf}
{\scshape ATLAS} collaboration, \emph{{Measurement of the angular coefficients
  in $Z$-boson events using electron and muon pairs from data taken at
  $\sqrt{s}=8$ TeV with the ATLAS detector}},
  \href{https://doi.org/10.1007/JHEP08(2016)159}{\emph{JHEP} {\bfseries 08}
  (2016) 159} [\href{https://arxiv.org/abs/1606.00689}{{\ttfamily
  1606.00689}}].

\bibitem{CMS:2015cyj}
{\scshape CMS} collaboration, \emph{{Angular coefficients of Z bosons produced
  in pp collisions at $\sqrt{s}$ = 8 TeV and decaying to $\mu^+ \mu^-$ as a
  function of transverse momentum and rapidity}},
  \href{https://doi.org/10.1016/j.physletb.2015.08.061}{\emph{Phys. Lett. B}
  {\bfseries 750} (2015) 154}
  [\href{https://arxiv.org/abs/1504.03512}{{\ttfamily 1504.03512}}].

\bibitem{LHCb:2022tbc}
{\scshape LHCb} collaboration, \emph{{First Measurement of the
  Z\textrightarrow{}\ensuremath{\mu}+\ensuremath{\mu}- Angular Coefficients in
  the Forward Region of pp Collisions at s=13~TeV}},
  \href{https://doi.org/10.1103/PhysRevLett.129.091801}{\emph{Phys. Rev. Lett.}
  {\bfseries 129} (2022) 091801}
  [\href{https://arxiv.org/abs/2203.01602}{{\ttfamily 2203.01602}}].

\bibitem{Karlberg:2014qua}
A.~Karlberg, E.~Re and G.~Zanderighi, \emph{{NNLOPS accurate Drell-Yan
  production}}, \href{https://doi.org/10.1007/JHEP09(2014)134}{\emph{JHEP}
  {\bfseries 09} (2014) 134} [\href{https://arxiv.org/abs/1407.2940}{{\ttfamily
  1407.2940}}].

\bibitem{Lambertsen:2016wgj}
M.~Lambertsen and W.~Vogelsang, \emph{{Drell-Yan lepton angular distributions
  in perturbative QCD}},
  \href{https://doi.org/10.1103/PhysRevD.93.114013}{\emph{Phys. Rev. D}
  {\bfseries 93} (2016) 114013}
  [\href{https://arxiv.org/abs/1605.02625}{{\ttfamily 1605.02625}}].

\bibitem{Gauld:2017tww}
R.~Gauld, A.~Gehrmann-De~Ridder, T.~Gehrmann, E.W.N.~Glover and A.~Huss,
  \emph{{Precise predictions for the angular coefficients in Z-boson production
  at the LHC}}, \href{https://doi.org/10.1007/JHEP11(2017)003}{\emph{JHEP}
  {\bfseries 11} (2017) 003}
  [\href{https://arxiv.org/abs/1708.00008}{{\ttfamily 1708.00008}}].

\bibitem{Gauld:2021pkr}
R.~Gauld, A.~Gehrmann-De~Ridder, T.~Gehrmann, E.W.N.~Glover, A.~Huss, I.~Majer
  et~al., \emph{{Transverse momentum distributions in low-mass Drell-Yan lepton
  pair production at NNLO QCD}},
  \href{https://doi.org/10.1016/j.physletb.2022.137111}{\emph{Phys. Lett. B}
  {\bfseries 829} (2022) 137111}
  [\href{https://arxiv.org/abs/2110.15839}{{\ttfamily 2110.15839}}].

\bibitem{Lyubovitskij:2024civ}
V.E.~Lyubovitskij, W.~Vogelsang, F.~Wunder and A.S.~Zhevlakov,
  \emph{{Perturbative T-odd asymmetries in the Drell-Yan process revisited}},
  \href{https://doi.org/10.1103/PhysRevD.109.114023}{\emph{Phys. Rev. D}
  {\bfseries 109} (2024) 114023}
  [\href{https://arxiv.org/abs/2403.18741}{{\ttfamily 2403.18741}}].

\bibitem{Lyubovitskij:2024jlb}
V.E.~Lyubovitskij, A.S.~Zhevlakov and I.A.~Anikin, \emph{{Transverse momentum
  dependence of the T-even hadronic structure functions in the Drell-Yan
  process}}, \href{https://doi.org/10.1103/PhysRevD.110.074028}{\emph{Phys.
  Rev. D} {\bfseries 110} (2024) 074028}
  [\href{https://arxiv.org/abs/2408.01243}{{\ttfamily 2408.01243}}].

\bibitem{Lyubovitskij:2025oig}
V.E.~Lyubovitskij, A.S.~Zhevlakov and I.A.~Anikin, \emph{{Angular coefficients
  of the Drell-Yan process across different rapidity and kinematical ranges}},
  \href{https://arxiv.org/abs/2503.16008}{{\ttfamily 2503.16008}}.

\bibitem{Barone:2010gk}
V.~Barone, S.~Melis and A.~Prokudin, \emph{{Azimuthal asymmetries in
  unpolarized Drell-Yan processes and the Boer-Mulders distributions of
  antiquarks}}, \href{https://doi.org/10.1103/PhysRevD.82.114025}{\emph{Phys.
  Rev. D} {\bfseries 82} (2010) 114025}
  [\href{https://arxiv.org/abs/1009.3423}{{\ttfamily 1009.3423}}].

\bibitem{Lu:2011mz}
Z.~Lu and I.~Schmidt, \emph{{The $\cos2\phi$ azimuthal asymmetry of unpolarized
  dilepton production at the $Z$-pole}},
  \href{https://doi.org/10.1103/PhysRevD.84.094002}{\emph{Phys. Rev. D}
  {\bfseries 84} (2011) 094002}
  [\href{https://arxiv.org/abs/1107.4693}{{\ttfamily 1107.4693}}].

\bibitem{Ebert:2020dfc}
M.A.~Ebert, J.K.L.~Michel, I.W.~Stewart and F.J.~Tackmann, \emph{{Drell-Yan
  $q_{T}$ resummation of fiducial power corrections at N$^{3}$LL}},
  \href{https://doi.org/10.1007/JHEP04(2021)102}{\emph{JHEP} {\bfseries 04}
  (2021) 102} [\href{https://arxiv.org/abs/2006.11382}{{\ttfamily
  2006.11382}}].

\bibitem{Piloneta:2024aac}
S.~Piloneta and A.~Vladimirov, \emph{{Angular distributions of Drell-Yan
  leptons in the TMD factorization approach}},
  \href{https://doi.org/10.1007/JHEP12(2024)059}{\emph{JHEP} {\bfseries 12}
  (2024) 059} [\href{https://arxiv.org/abs/2407.06277}{{\ttfamily
  2407.06277}}].

\bibitem{Collins:1977iv}
J.C.~Collins and D.E.~Soper, \emph{{Angular Distribution of Dileptons in
  High-Energy Hadron Collisions}},
  \href{https://doi.org/10.1103/PhysRevD.16.2219}{\emph{Phys. Rev. D}
  {\bfseries 16} (1977) 2219}.

\bibitem{Mirkes:1994eb}
E.~Mirkes and J.~Ohnemus, \emph{{$W$ and $Z$ polarization effects in hadronic
  collisions}}, \href{https://doi.org/10.1103/PhysRevD.50.5692}{\emph{Phys.
  Rev. D} {\bfseries 50} (1994) 5692}
  [\href{https://arxiv.org/abs/hep-ph/9406381}{{\ttfamily hep-ph/9406381}}].

\bibitem{Aybat:2011zv}
S.M.~Aybat and T.C.~Rogers, \emph{{TMD Parton Distribution and Fragmentation
  Functions with QCD Evolution}},
  \href{https://doi.org/10.1103/PhysRevD.83.114042}{\emph{Phys. Rev. D}
  {\bfseries 83} (2011) 114042}
  [\href{https://arxiv.org/abs/1101.5057}{{\ttfamily 1101.5057}}].

\bibitem{Chiu:2012ir}
J.-Y.~Chiu, A.~Jain, D.~Neill and I.Z.~Rothstein, \emph{{A Formalism for the
  Systematic Treatment of Rapidity Logarithms in Quantum Field Theory}},
  \href{https://doi.org/10.1007/JHEP05(2012)084}{\emph{JHEP} {\bfseries 05}
  (2012) 084} [\href{https://arxiv.org/abs/1202.0814}{{\ttfamily 1202.0814}}].

\bibitem{Scimemi:2018xaf}
I.~Scimemi and A.~Vladimirov, \emph{{Systematic analysis of double-scale
  evolution}}, \href{https://doi.org/10.1007/JHEP08(2018)003}{\emph{JHEP}
  {\bfseries 08} (2018) 003}
  [\href{https://arxiv.org/abs/1803.11089}{{\ttfamily 1803.11089}}].

\bibitem{Lee:2022nhh}
R.N.~Lee, A.~von Manteuffel, R.M.~Schabinger, A.V.~Smirnov, V.A.~Smirnov and
  M.~Steinhauser, \emph{{Quark and Gluon Form Factors in Four-Loop QCD}},
  \href{https://doi.org/10.1103/PhysRevLett.128.212002}{\emph{Phys. Rev. Lett.}
  {\bfseries 128} (2022) 212002}
  [\href{https://arxiv.org/abs/2202.04660}{{\ttfamily 2202.04660}}].

\bibitem{Duhr:2022yyp}
C.~Duhr, B.~Mistlberger and G.~Vita, \emph{{Four-Loop Rapidity Anomalous
  Dimension and Event Shapes to Fourth Logarithmic Order}},
  \href{https://doi.org/10.1103/PhysRevLett.129.162001}{\emph{Phys. Rev. Lett.}
  {\bfseries 129} (2022) 162001}
  [\href{https://arxiv.org/abs/2205.02242}{{\ttfamily 2205.02242}}].

\bibitem{Moult:2022xzt}
I.~Moult, H.X.~Zhu and Y.J.~Zhu, \emph{{The four loop QCD rapidity anomalous
  dimension}}, \href{https://doi.org/10.1007/JHEP08(2022)280}{\emph{JHEP}
  {\bfseries 08} (2022) 280}
  [\href{https://arxiv.org/abs/2205.02249}{{\ttfamily 2205.02249}}].

\bibitem{Manohar:2002fd}
A.V.~Manohar, T.~Mehen, D.~Pirjol and I.W.~Stewart, \emph{{Reparameterization
  invariance for collinear operators}},
  \href{https://doi.org/10.1016/S0370-2693(02)02029-4}{\emph{Phys. Lett. B}
  {\bfseries 539} (2002) 59}
  [\href{https://arxiv.org/abs/hep-ph/0204229}{{\ttfamily hep-ph/0204229}}].

\bibitem{Marcantonini:2008qn}
C.~Marcantonini and I.W.~Stewart, \emph{{Reparameterization Invariant Collinear
  Operators}}, \href{https://doi.org/10.1103/PhysRevD.79.065028}{\emph{Phys.
  Rev. D} {\bfseries 79} (2009) 065028}
  [\href{https://arxiv.org/abs/0809.1093}{{\ttfamily 0809.1093}}].

\bibitem{Mulders:1995dh}
P.J.~Mulders and R.D.~Tangerman, \emph{{The Complete tree level result up to
  order 1/Q for polarized deep inelastic leptoproduction}},
  \href{https://doi.org/10.1016/0550-3213(95)00632-X}{\emph{Nucl. Phys. B}
  {\bfseries 461} (1996) 197}
  [\href{https://arxiv.org/abs/hep-ph/9510301}{{\ttfamily hep-ph/9510301}}].

\bibitem{Bacchetta:2006tn}
A.~Bacchetta, M.~Diehl, K.~Goeke, A.~Metz, P.J.~Mulders and M.~Schlegel,
  \emph{{Semi-inclusive deep inelastic scattering at small transverse
  momentum}}, \href{https://doi.org/10.1088/1126-6708/2007/02/093}{\emph{JHEP}
  {\bfseries 02} (2007) 093}
  [\href{https://arxiv.org/abs/hep-ph/0611265}{{\ttfamily hep-ph/0611265}}].

\bibitem{Moos:2020wvd}
V.~Moos and A.~Vladimirov, \emph{{Calculation of transverse momentum dependent
  distributions beyond the leading power}},
  \href{https://doi.org/10.1007/JHEP12(2020)145}{\emph{JHEP} {\bfseries 12}
  (2020) 145} [\href{https://arxiv.org/abs/2008.01744}{{\ttfamily
  2008.01744}}].

\bibitem{Kanazawa:2015ajw}
K.~Kanazawa, Y.~Koike, A.~Metz, D.~Pitonyak and M.~Schlegel, \emph{{Operator
  Constraints for Twist-3 Functions and Lorentz Invariance Properties of
  Twist-3 Observables}},
  \href{https://doi.org/10.1103/PhysRevD.93.054024}{\emph{Phys. Rev. D}
  {\bfseries 93} (2016) 054024}
  [\href{https://arxiv.org/abs/1512.07233}{{\ttfamily 1512.07233}}].

\bibitem{Bacchetta:2013pqa}
A.~Bacchetta and A.~Prokudin, \emph{{Evolution of the helicity and transversity
  Transverse-Momentum-Dependent parton distributions}},
  \href{https://doi.org/10.1016/j.nuclphysb.2013.07.013}{\emph{Nucl. Phys. B}
  {\bfseries 875} (2013) 536}
  [\href{https://arxiv.org/abs/1303.2129}{{\ttfamily 1303.2129}}].

\bibitem{Dai:2014ala}
L.-Y.~Dai, Z.-B.~Kang, A.~Prokudin and I.~Vitev, \emph{{Next-to-leading order
  transverse momentum-weighted Sivers asymmetry in semi-inclusive deep
  inelastic scattering: the role of the three-gluon correlator}},
  \href{https://doi.org/10.1103/PhysRevD.92.114024}{\emph{Phys. Rev. D}
  {\bfseries 92} (2015) 114024}
  [\href{https://arxiv.org/abs/1409.5851}{{\ttfamily 1409.5851}}].

\bibitem{Echevarria:2016scs}
M.G.~Echevarria, I.~Scimemi and A.~Vladimirov, \emph{{Unpolarized Transverse
  Momentum Dependent Parton Distribution and Fragmentation Functions at
  next-to-next-to-leading order}},
  \href{https://doi.org/10.1007/JHEP09(2016)004}{\emph{JHEP} {\bfseries 09}
  (2016) 004} [\href{https://arxiv.org/abs/1604.07869}{{\ttfamily
  1604.07869}}].

\bibitem{Gutierrez-Reyes:2018iod}
D.~Gutierrez-Reyes, I.~Scimemi and A.~Vladimirov, \emph{{Transverse momentum
  dependent transversely polarized distributions at
  next-to-next-to-leading-order}},
  \href{https://doi.org/10.1007/JHEP07(2018)172}{\emph{JHEP} {\bfseries 07}
  (2018) 172} [\href{https://arxiv.org/abs/1805.07243}{{\ttfamily
  1805.07243}}].

\bibitem{Gutierrez-Reyes:2019rug}
D.~Gutierrez-Reyes, S.~Leal-Gomez, I.~Scimemi and A.~Vladimirov,
  \emph{{Linearly polarized gluons at next-to-next-to leading order and the
  Higgs transverse momentum distribution}},
  \href{https://doi.org/10.1007/JHEP11(2019)121}{\emph{JHEP} {\bfseries 11}
  (2019) 121} [\href{https://arxiv.org/abs/1907.03780}{{\ttfamily
  1907.03780}}].

\bibitem{Scimemi:2019gge}
I.~Scimemi, A.~Tarasov and A.~Vladimirov, \emph{{Collinear matching for Sivers
  function at next-to-leading order}},
  \href{https://doi.org/10.1007/JHEP05(2019)125}{\emph{JHEP} {\bfseries 05}
  (2019) 125} [\href{https://arxiv.org/abs/1901.04519}{{\ttfamily
  1901.04519}}].

\bibitem{Ebert:2020qef}
M.A.~Ebert, B.~Mistlberger and G.~Vita, \emph{{TMD Fragmentation Functions at
  N$^3$LO}}, \href{https://doi.org/10.1007/JHEP07(2021)121}{\emph{JHEP}
  {\bfseries 07} (2021) 121}
  [\href{https://arxiv.org/abs/2012.07853}{{\ttfamily 2012.07853}}].

\bibitem{Luo:2020epw}
M.-x.~Luo, T.-Z.~Yang, H.X.~Zhu and Y.J.~Zhu, \emph{{Unpolarized quark and
  gluon TMD PDFs and FFs at N$^{3}$LO}},
  \href{https://doi.org/10.1007/JHEP06(2021)115}{\emph{JHEP} {\bfseries 06}
  (2021) 115} [\href{https://arxiv.org/abs/2012.03256}{{\ttfamily
  2012.03256}}].

\bibitem{Ebert:2020yqt}
M.A.~Ebert, B.~Mistlberger and G.~Vita, \emph{{Transverse momentum dependent
  PDFs at N$^3$LO}}, \href{https://doi.org/10.1007/JHEP09(2020)146}{\emph{JHEP}
  {\bfseries 09} (2020) 146}
  [\href{https://arxiv.org/abs/2006.05329}{{\ttfamily 2006.05329}}].

\bibitem{Rein:2022odl}
F.~Rein, S.~Rodini, A.~Sch\"afer and A.~Vladimirov, \emph{{Sivers, Boer-Mulders
  and worm-gear distributions at next-to-leading order}},
  \href{https://doi.org/10.1007/JHEP01(2023)116}{\emph{JHEP} {\bfseries 01}
  (2023) 116} [\href{https://arxiv.org/abs/2209.00962}{{\ttfamily
  2209.00962}}].

\bibitem{Echevarria:2015uaa}
M.G.~Echevarria, T.~Kasemets, P.J.~Mulders and C.~Pisano, \emph{{QCD evolution
  of (un)polarized gluon TMDPDFs and the Higgs $q_T$-distribution}},
  \href{https://doi.org/10.1007/JHEP07(2015)158}{\emph{JHEP} {\bfseries 07}
  (2015) 158} [\href{https://arxiv.org/abs/1502.05354}{{\ttfamily
  1502.05354}}].

\bibitem{Scimemi:2018mmi}
I.~Scimemi and A.~Vladimirov, \emph{{Matching of transverse momentum dependent
  distributions at twist-3}},
  \href{https://doi.org/10.1140/epjc/s10052-018-6263-5}{\emph{Eur. Phys. J. C}
  {\bfseries 78} (2018) 802}
  [\href{https://arxiv.org/abs/1804.08148}{{\ttfamily 1804.08148}}].

\bibitem{Boer:2006eq}
D.~Boer and W.~Vogelsang, \emph{{Drell-Yan lepton angular distribution at small
  transverse momentum}},
  \href{https://doi.org/10.1103/PhysRevD.74.014004}{\emph{Phys. Rev. D}
  {\bfseries 74} (2006) 014004}
  [\href{https://arxiv.org/abs/hep-ph/0604177}{{\ttfamily hep-ph/0604177}}].

\bibitem{Moos:2023yfa}
V.~Moos, I.~Scimemi, A.~Vladimirov and P.~Zurita, \emph{{Extraction of
  unpolarized transverse momentum distributions from the fit of Drell-Yan data
  at N$^{4}$LL}}, \href{https://doi.org/10.1007/JHEP05(2024)036}{\emph{JHEP}
  {\bfseries 05} (2024) 036}
  [\href{https://arxiv.org/abs/2305.07473}{{\ttfamily 2305.07473}}].

\end{thebibliography}\endgroup
